\documentclass[review,11pt]{elsarticle}

\usepackage{amssymb}
\usepackage{bm}
\usepackage{lineno,hyperref}
\usepackage{miller}
\usepackage{color}
\usepackage{ulem}

\journal{Acta Materialia}

\begin{document}

\begin{frontmatter}



\title{The structure and migration of heavily irradiated grain boundaries and dislocations in Ni in the athermal limit}

\author[XCP5]{Ian Chesser\corref{mycorrespondingauthor}}
\author[PSI]{Peter M. Derlet}
\author[T1]{Avanish Mishra}
\author[MPA]{Sarah Paguaga}
\author[T1]{Nithin Mathew}
\author[MST]{Khanh Dang}
\author[MST]{Blas Pedro Uberuaga}
\author[XCP5]{Abigail Hunter}
\author[MPA]{Saryu Fensin}

\cortext[mycorrespondingauthor]{Corresponding author}

\address[XCP5]{XCP-5, Los Alamos National Laboratory, Los Alamos, New Mexico, NM USA}
\address[PSI]{Condensed Matter Theory Group, Paul Scherrer Institute, Villigen PSI, Switzerland}
\address[T1]{T-1, Los Alamos National Laboratory, Los Alamos, New Mexico, NM USA}
\address[MPA]{MPA-CINT, Los Alamos National Laboratory, Los Alamos, New Mexico, NM USA}
\address[MST]{MST-8, Los Alamos National Laboratory, Los Alamos, New Mexico, NM USA}


%

\begin{abstract}

The microstructural evolution at and near pre-existing grain boundaries (GBs) and dislocations in materials under high radiation doses is still poorly understood. In this work, we use the creation relaxation algorithm (CRA) developed for atomistic modeling of high-dose irradiation in bulk materials to probe the athermal limit of saturation of GB and dislocation core regions under irradiation in FCC Ni. We find that, upon continuously subjecting a single dislocation or GB to Frenkel pair creation in the athermal limit, a local steady state disordered defect structure is reached with excess properties that fluctuate around constant values. Case studies are given for a straight screw dislocation which elongates into a helix under irradiation and several types of low and high angle GBs, which exhibit coupled responses such as absorption of extrinsic dislocations, roughening and migration. A positive correlation is found between initial GB energy and the local steady state GB energy under irradiation across a wide variety of GB types. Metastable GB structures with similar density in the defect core region but different initial configurations are found to converge to the same limiting structure under CRA. The mechanical responses of pristine and irradiated dislocations and GB structures are compared under an applied shear stress. Irradiated screw and edge dislocations are found to exhibit a hardening response, migrating at larger flow stresses than their pristine counterparts. Mobile GBs are found to exhibit softening or hardening responses depending on GB character. Although some GBs recover their initial pristine structures upon migration outside of the radiation zone, many GBs sustain different flow stresses corresponding to altered mobile core structures.

\end{abstract}

\end{frontmatter}


\section{Introduction}
\label{intro}

Increasing the pre-existing dislocation and grain boundary (GB) density in material microstructures is one way to improve radiation damage tolerance of crystalline materials because these topological defects act as storage sites, or sinks, for radiation induced defects such as interstitials, vacancies and prismatic dislocation loops \cite{zinkle20121,beyerlein2015defect,zhang2018radiation,barr2019interplay,kohnert2019sink}. However, GBs and dislocations are highly susceptible to radiation induced migration, leading to deleterious microstructural instabilities, such as grain growth in nano-crystalline materials, that reduce material strength and radiation damage tolerance \cite{cunningham2021suppressing,stangebye2023direct,thomas2023thermal}. There is an ongoing effort to design materials with tailored compositions and defect distributions that have simultaneous radiation tolerance and microstructural stability up to large doses $>$ 100 displacements per atom (DPA) commonly reached in extreme environments. 

The sink action of topological defects is still only partially understood, especially at large radiation doses. Accommodation of radiation damage by GBs and dislocations is a complex process involving defect reaction and diffusion processes at multiple time and length scales. There is a consensus that topological defect character has a strong impact on radiation damage accumulation. Some special GBs such as the coherent twin in face centered cubic (FCC) metals behave as ineffective sinks which enhance damage in the surrounding grains, while other GBs behave as nearly perfect sinks which continuously annihilate radiation damage \cite{sutton1995interfaces, han2012effect, bai2013influence, el2017direct}. It is recognized that GBs are not static sinks for point defects but rather have sink efficiencies which vary with GB structure and character, and irradiation conditions such as dose, dose rate, ion type, and temperature \cite{sutton1995interfaces, uberuaga2015relationship, el2018does, el2024grain}. Taheri et al. found that non-equilibrium GB structures quenched from an irradiated state have a higher average sink efficiency than well-annealed GBs \cite{NEGB_expt}. Non-equilibrium rejuvenated GB structures formed by fast laser pulses were found to be softer on average than well-annealed GBs, representing a tradeoff between radiation damage tolerance and mechanical strength \cite{balbus2018femtosecond}. The dependence of radiation damage tolerance on GB metastability is a complicating factor for modeling trends of radiation damage with topological defect character. 

Atomic-scale simulations allow for the most detailed account of radiation damage mechanisms at GBs and dislocations to date, but have traditionally been performed at doses $<$ 0.1 DPA, smaller than typical irradiation experiments or service conditions. It is challenging to study large doses in atomistic simulations because of the long times required to simulate individual collision cascades and thermal evolution between cascades. Only a handful of simulations have considered the evolution of GB structure under continuous irradiation damage at doses O(0.1 DPA) and higher \cite{yu2015non, NEGB_sim2, jiang2017multiple, jin2018radiation}. An important result from these works is that, after sufficient damage, at least some GBs show signs of structural saturation, suggesting a limiting behavior that is useful for examining trends of radiation damage across many GBs. Cascade simulations in two Cu bicrystals with symmetric tilt GBs irradiated up to a total dose of 0.3 DPA revealed saturation of stacking fault tetrahedron concentrations at the GBs by 0.1 DPA \cite{jin2018radiation}. Low angle twist GBs in SiC and high angle tilt GBs in Cu continuously loaded by interstitials and vacancies exhibited periodic variation in GB energies corresponding to dislocation climb behavior in the plane of the GB in the case of twist GBs and net GB migration in the case of tilt GBs \cite{jiang2017multiple}. Tucker et al. showed that the excess energy and volume of a variety of high angle Cu tilt GB structures saturate under continuous vacancy loading, finding a correlation between initial GB energy, the total excess energy and the excess volume of saturated non-equilibrium GB structures \cite{NEGB_sim2}. Compared to GBs, even less work has been done to study radiation damage near dislocations at large doses. We are not aware of any simulations studying the irradiation-induced saturation of local microstructure in the vicinity of pre-existing dislocations. The existence of steady state behavior under irradiation in the vicinity of topological defects has yet to be verified for a wide range of GBs and dislocations over a large range of doses. 

In a separate line of work, several recent simulation studies have examined microstructural evolution in bulk materials in the limit of athermal accumulation of Frenkel pairs (FPs) \cite{CRA_BCC, CRA_glass, ma2023athermal}. The Creation Relaxation Algorithm (CRA) consists of iterative FP insertion and energy minimization and bypasses the costly simulation of individual collision cascades, allowing for the simulation of relatively large radiation doses O(10 DPA) in the athermal limit of infinite radiation dose rate \cite{CRA_BCC, granberg2023effect}. A more detailed discussion of the advantages and shortcomings of the CRA method is given in Section \ref{cra}. Because of the large total dose, the CRA method allows for the identification of possible radiation-induced steady state microstructures that arise from collective stress driven defect behavior. Close-packed metals and certain oxide materials have been found to exhibit the formation of system spanning dislocation networks at steady state with regimes of swelling consistent with experiments \cite{mason2020observation, chartier2016early}. Well-relaxed glass structures were found to exhibit steady state rejuvenated structures under CRA with system-spanning backbones of icosahedrally coordinated atoms broken up by the radiation damage \cite{CRA_glass}. Application of CRA to nano-crystalline tungsten resulted in irradiation-induced grain growth and the formation of vacancy and dislocation denuded zones near GBs \cite{ma2023athermal}. In the case of grain growth, no steady state was reached by a total dose of 3 DPA, demonstrating that an existing topological defect network can delay the onset of a steady state compared to a perfect crystal. 

In this work, we apply CRA to dislocation and GB core regions in FCC Ni at doses up to 20 DPA and find that local microstructure in the vicinity of pre-existing GBs and dislocations saturates at sufficiently large radiation doses for a wide range of GB and dislocation types. We demonstrate that irradiation induced steady state local microstructures often have different yield and migration behavior under an applied shear strain compared to their pristine counterparts, depending on GB and dislocation character. We characterize microstructural evolution processes at an approaching steady state and analyze trends in damage accumulation behavior with GB and dislocation character, finding that high energy pristine GBs saturate more easily  than low energy GBs. 

\section{Methods}
\label{methods}
\noindent The Large-scale Atomic/Molecular Massively Parallel Simulator
(LAMMPS) \citep{LAMMPS} was utilized to conduct Molecular Statics (MS) and dynamics (MD) simulations. The software package OVITO
\citep{OVITO} was used to visualize and analyze GB structures
and their radiation damage responses. Dislocation character was quantified via the DXA algorithm \cite{DXA} in OVITO. Local atomic environments were visualized using the Polyhedral Template Matching (PTM) algorithm \cite{larsen2016robust} in OVITO with local structures identified as FCC = Face Centered Cubic, HCP = Hexagonal Close Packed, BCC = Body Centered Cubic, ICO = Icosahedral, and Other = disordered. Interstitial and vacancy sites were analyzed via Wigner-Seitz analysis in OVITO using pristine un-irradiated samples as reference structures. We note that Wigner-Seitz analysis allows for both vacancy and interstitial positions to be computed within the reference pristine structure, but only for interstitial positions to be computed within the current irradiated structure. 

\subsection{The philosophy of the CRA approach}
\label{cra}

While finite temperature atomistic simulation of cascade evolution has provided great insight into the very earliest stages of irradiation --- the primary damage state \cite{de1991new} --- it is unable to model the long times associated with thermally driven (but stress biased) defect transport and reaction, and more generally, the microstructural evolution that is experimentally known to occur as the dose increases. To do this, multi-scale modeling strategies are needed that begin from the dilute defect limit of the primary damage state, where microstructural fluctuation (defect migration) timescales are faster than the timescale between cascades, the inverse dose-rate.  As already discussed, experiments demonstrate the formation of complex microstructure under heavy irradiation, involving (in part) system-spanning dislocation networks. This microstructural regime is far from the dilute limit and has a relevant fluctuation timescale significantly longer than the inverse dose-rate, motivating an alternative modeling strategy in which a defect structure has little time to relax before the next cascade. The limiting form of this perspective is the absence of activated processes, with the material response being entirely athermal, and thus stress driven and at zero temperature. 

The CRA method \cite{CRA_BCC} is the simplest realization of such an approach, producing a system spanning microstructure whose statistical properties may converge to a steady-state structure at a sufficiently high dose. While such CRA produced microstructures have qualitative similarities to what is seen in experiment \cite{mason2020observation}, it is natural to ask how quantitatively similar they are to microstructures observed in the experimental domain of finite temperature and lower dose-rates.  Past multi-cascade simulations at finite temperature \cite{byggmastar2018effects} reveal a qualitatively similar evolution to that seen in CRA \cite{CRA_BCC}, an expected result, since the timescales of any finite temperature atomistic simulation will access only the fastest thermally driven processes and involve dose rates many orders of magnitude faster than that of an experiment. The quantitative relevance of CRA produced structures remains an open question. The strength of the CRA approach is that it provides a well defined asymptote from which one can begin to model irradiated microstructure in the large dose limit. 

An additional aspect is that any CRA derived steady state will always be dependent on the chosen boundary conditions, such as the finite size of the simulation cell, whether or not they are open or periodic, and the nature of the imposed stress/strain state. For example, when applying CRA to a perfect lattice under full periodic boundary conditions, the topological criterion of zero total Burgers vector content sets an upper limit for the effective screening length of dislocations to the system size. If the CRA irradiation zone  differs from the full simulation cell, there will be an additional dependence on its size and shape. This is the case for the present work since finite irradiation regions are centered on linear and planar defects of interest. Any referred steady state is implicitly local to the irradiated region. The motivation for the present approach is an interest in how such topological defects themselves respond to the CRA procedure, rather than how these defects and their micro-structural environments respond together. 

\subsection{GB dataset} 
\label{gbdataset}

Ni was chosen as a model FCC system to study irradiation damage at GBs with interatomic interactions parametrized by the Foiles-Hoyt Ni EAM potential \cite{foiles2006computation}. This choice was motivated by the availability of a well-studied atomistic database of 388 Ni GBs with diverse macroscopic GB crystallography \cite{olmsted2009survey}. Types of GBs in the dataset include symmetric tilt and twist GBs as well as asymmetric tilt GBs and GBs with general mixed tilt/twist character. For a given misorientation, multiple boundary plane inclinations are sampled \cite{homer2015grain}. Two system sizes were considered in this work corresponding to replicated GB plane sizes of approximately (3 nm x 3 nm) and (10 nm x 10 nm). In both cases, grain sizes were at least 10 nm normal to the GB plane. Previous studies of this dataset have focused on GB energy, mobility, shear coupling factor, migration mechanisms, and GB-dislocation interactions \cite{olmsted2009survey,olmsted2009survey2,yu2021survey,chesser2022taxonomy,adams2019atomistic}. 

\subsection{Dislocation dataset} 
\label{ddataset}

Initial unrelaxed dislocation structures in Ni were generated with the atomman Python package \cite{atomman}. Individual perfect dislocations with pure screw and edge character were created in a rectangular system with periodic boundary conditions along the dislocation line in the X direction and free surface boundary conditions normal to the dislocation line. To create each dislocation, a displacement field was applied consistent with the solution from anisotropic linear elasticity theory for a straight screw or edge dislocation with Burgers vector $\frac{1}{2}\hkl<110>$ using the elastic constants for the chosen Ni potential. In the case of the screw dislocation, the X axis was parallel to a \hkl<110> direction and the Y and Z axes were parallel to \hkl<112> and \hkl<111> directions. In the case of the edge dislocation, the X axis was parallel to a \hkl<112> direction, whereas the Y and Z axes were parallel to \hkl<110> and \hkl<111> directions. The total box size was chosen to be approximately 20 nm along the dislocation line and 10 nm normal to the dislocation line.

\subsection{Metastable structure generation via quench from a glassy state}

\begin{figure}[ht!]
\centering\leavevmode \includegraphics[width=0.99\textwidth]{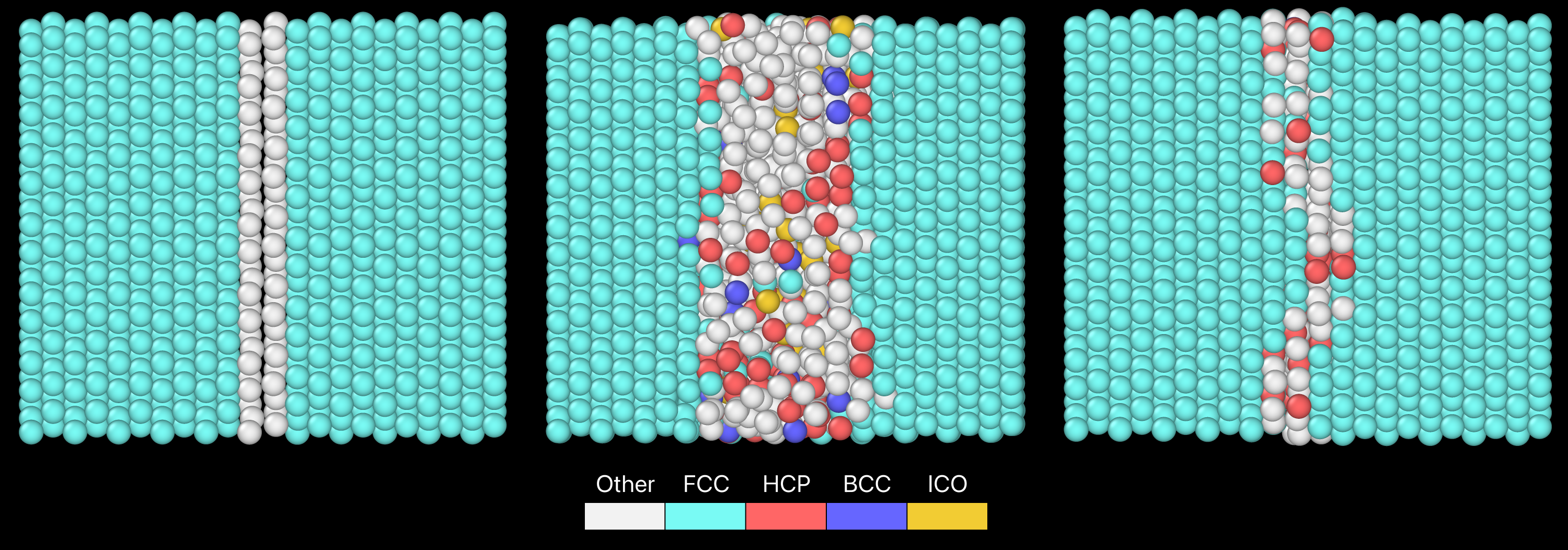}
\caption{Method to create metastable GB structures illustrated with the $\Sigma 5$ \hkl{100} twist GB. (a) Pristine structure is hollowed out and replaced by (b) randomly deposited atoms with number equal to the initial structure (c) Upon quenching to 0 K, the disordered structure crystallizes with quenched-in defects. The local atomic environments shown in color bar come from the PTM modifier in OVITO.}
\label{fig:meta1}
\end{figure}

High energy metastable GB structures were generated from minimum energy GB structures via a random solidification protocol illustrated in Fig. \ref{fig:meta1}. First, the low energy GB core was removed by carving out a vacuum region with a thickness of 1.5 nm spanning the original GB. Next, atoms were randomly deposited into the vacuum region at T = 0.1 K at a rate of 100 atoms/ps. Bulk crystals were allowed to translate during this process, relieving high stresses in the GB normal direction that could build up during the deposition and allowing for possible GB sliding events. After a target number of deposited atoms was reached, structures were quenched to 0 K via a conjugate gradient minimization. Partial crystallization of the disordered GB was observed in this step with a resultant high energy GB structure with quenched-in defects. By changing the number of deposited atoms, this method can sample a wide range of GB densities. In this work, we primarily focus on the mass conserving case where the same number of atoms is deposited as was initially deleted. 

\subsection{Radiation damage protocol} 
\label{protocol}

\begin{figure}[ht!]
\centering\leavevmode \includegraphics[width=0.9\textwidth]{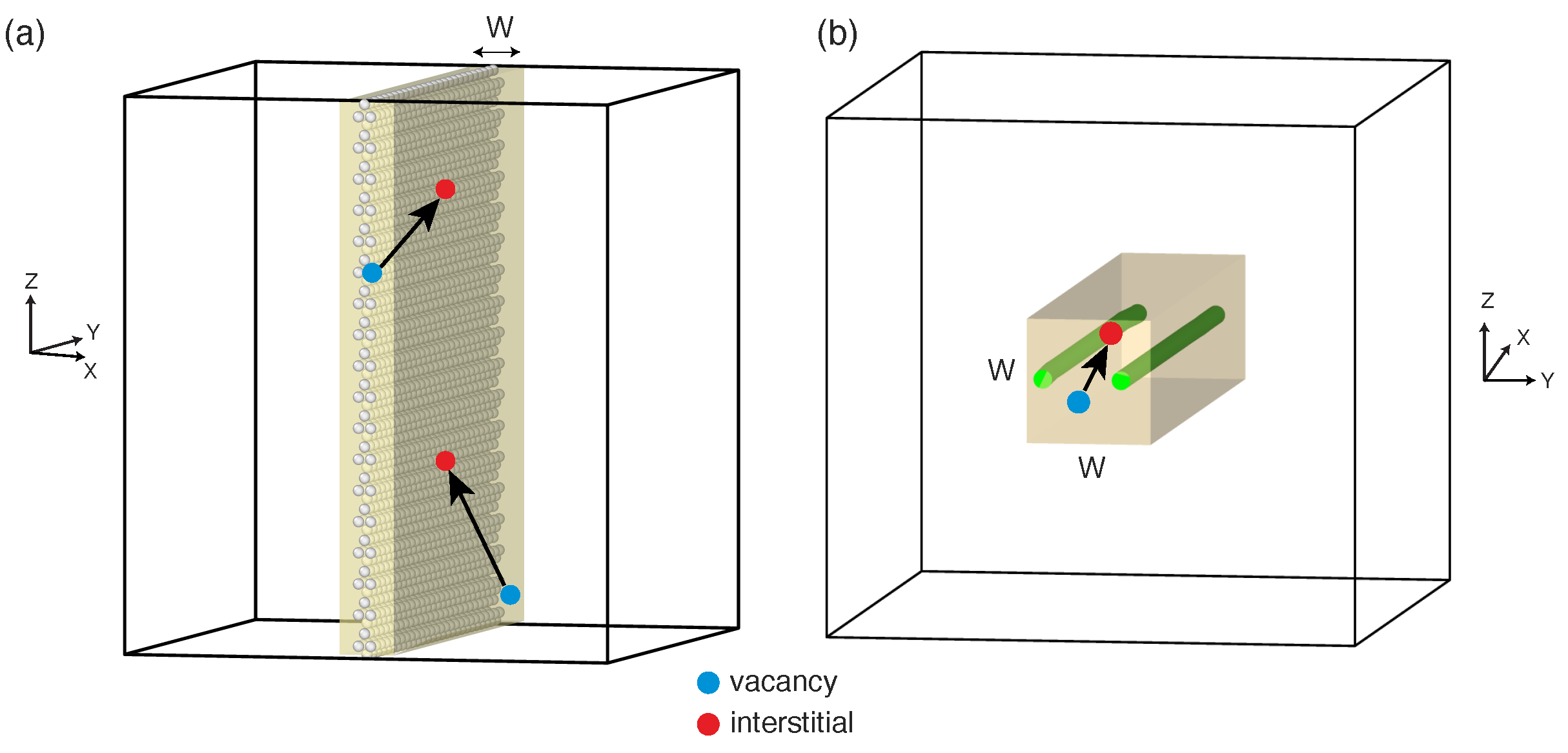}
\caption{Schematic of the CRA method. Vacancies and interstitials are generated within a finite radiation zone (yellow box) spanning the (a) GB and (b) dislocation core regions. Non-FCC atoms are shown for the GB. Partial dislocation lines are shown as segmented by DXA.}
\label{fig:FPI}
\end{figure}

The CRA algorithm was applied to a radiation zone of finite width $W$ centered at and spanning the GB core region of each bi-crystal, as shown in Fig. \ref{fig:FPI}. The zone width was set to 1 nm for the GB simulations in this work unless explicitly stated otherwise. Such a small zone width is intended to probe the limit of the GB core response in which point defects are attracted to the GB core. The radiation damage protocol consisted of iteratively inserting FPs in the radiation zone and relaxing the system via a conjugate gradient minimization at 0 K with force tolerance of $10^{-8}$ eV/ {\AA} and dimensionless energy tolerance $10^{-8}$. FP insertion involved randomly displacing atoms within the radiation zone (with no additional constraints on the magnitude or direction of displacements). To accelerate convergence to a steady state, multiple FPs ($N_{FP}$) were inserted prior to each minimization step. For a given radiation zone size $W$ and GB plane size, the steady state energy was found to be insensitive to $N_{FP}$ over the range $N_{FP}=1-60$ employed in this work. Radiation dose was measured in canonical DPA (cDPA), defined as the total number of FPs inserted divided by the total number of atoms in the radiation zone. The word canonical distinguishes the simple definition of irradiation dose in this work from definitions of DPA which may assume certain parameters tied to a particular irradiation source or collision cascade model. Periodic boundary conditions were employed normal to the GB plane, emulating a constrained polycrystalline setting and limiting the extent of GB sliding under irradiation. 

CRA was applied to all 388 Ni GBs for two GB plane sizes up to 50 cDPA for small GBs (3 nm x 3 nm plane size) and 20 cDPA for large GBs (10 nm x 10 nm plane size) in a narrow 1 nm radiation zone spanning the GB core regions. This large dataset allows for the statistical analysis of steady state energies and dynamics and the extraction of correlations which are applicable to many types of GBs, both special and general. Steady state properties such as excess energy were averaged over dose ranges starting at 5 cDPA and continuing up to the highest applied dose. A minority of GBs (around 20 GBs) in the dataset were filtered from the analysis because they did not meet the steady state criterion of a small standard deviation in energy fluctuations relative to the mean. These boundaries typically underwent significant migration with concomitant jumps in energy at late doses arising from the accumulation of bulk defect content in the radiation zone behind the mobile GB. We attempt to capture trends in the energetics of point defect accumulation in the core region of stationary GBs. 

Dislocation irradiation simulations employed a radiation zone defined as a ($L\times W\times W$) rectangular prism centered at and surrounding each dislocation line, where $L$ is the periodic dimension along the dislocation line and $W$ is the finite width of the radiation zone. $L$ was chosen to be 20 nm and $W$ was chosen to be 2 nm in order to fully span the dissociated dislocation core. The FP insertion procedure inside the radiation zone was analogous to the GB case. Minimization was performed in a cylindrical region around the dislocation core with a diameter of 7 nm, freezing all surface atoms.

\subsection{Strain driven migration of grain boundaries and dislocations} 

To simulate GB migration, a pure shear strain was applied to each bicrystal by displacing fixed slabs at the top and bottom of the bicrystal in opposite directions along the Z direction in the plane of the boundary. The Z direction is perpendicular to the tilt axis of symmetric tilt GBs and parallel to the Burgers vectors of their component edge dislocations. Slab displacements in steps of 0.04 {\AA} per slab were accompanied by a conjugate gradient energy minimization at 0 K with force tolerance of $10^{-8}$ eV/{\AA} and dimensionless energy tolerance $10^{-8}$. All bicrystals underwent 1000 displacement steps, or 8 nm total shear displacement corresponding to a shear strain of at least 0.2 for each GB. At sufficiently large shear strains, all GBs deformed plastically. GBs exhibited a diverse array of plastic deformation mechanisms including shear coupled motion, GB sliding, partial dislocation emission, and combinations thereof. 

A similar simulation set-up was used to drive dislocation motion via the application of a shear strain. Dislocations were constructed following \cite{dang2018pressure} with free surface boundary conditions along one dimension and periodic boundary conditions along the other two. Relaxed dislocation structures at 0 K were verified to be identical compared to dislocations produced with the cylindrical geometry discussed previously. Slabs defined at the free surfaces were displaced parallel to the Burgers vector of the edge dislocation and screw dislocation at a strain rate of $10^{8} $ s$^{-1}$  at 300 K. The entire simulation block was uniformly strained by a pre-computed thermal expansion coefficient for Ni at 300 K with deformation simulations performed in the NVT ensemble. A finite temperature was used to avoid twin growth during the deformation of the irradiated screw dislocations. In some cases, additional runs were performed with stress control as in \cite{dang2019mobility} to verify mechanisms observed under strain control. Simulation blocks contained approximately 1.2 million atoms with dimensions 20 nm along the dislocation line direction, 50 nm along the direction of dislocation motion, and 20 nm along the non-periodic direction. For irradiated samples, identical radiation zone sizes and FP block sizes were chosen compared to the cylindrical geometry.

\section{Results}
\label{results}

\subsection{Irradiation response of dislocations}

\begin{figure}[ht!]
\centering\leavevmode \includegraphics[width=0.99\textwidth]{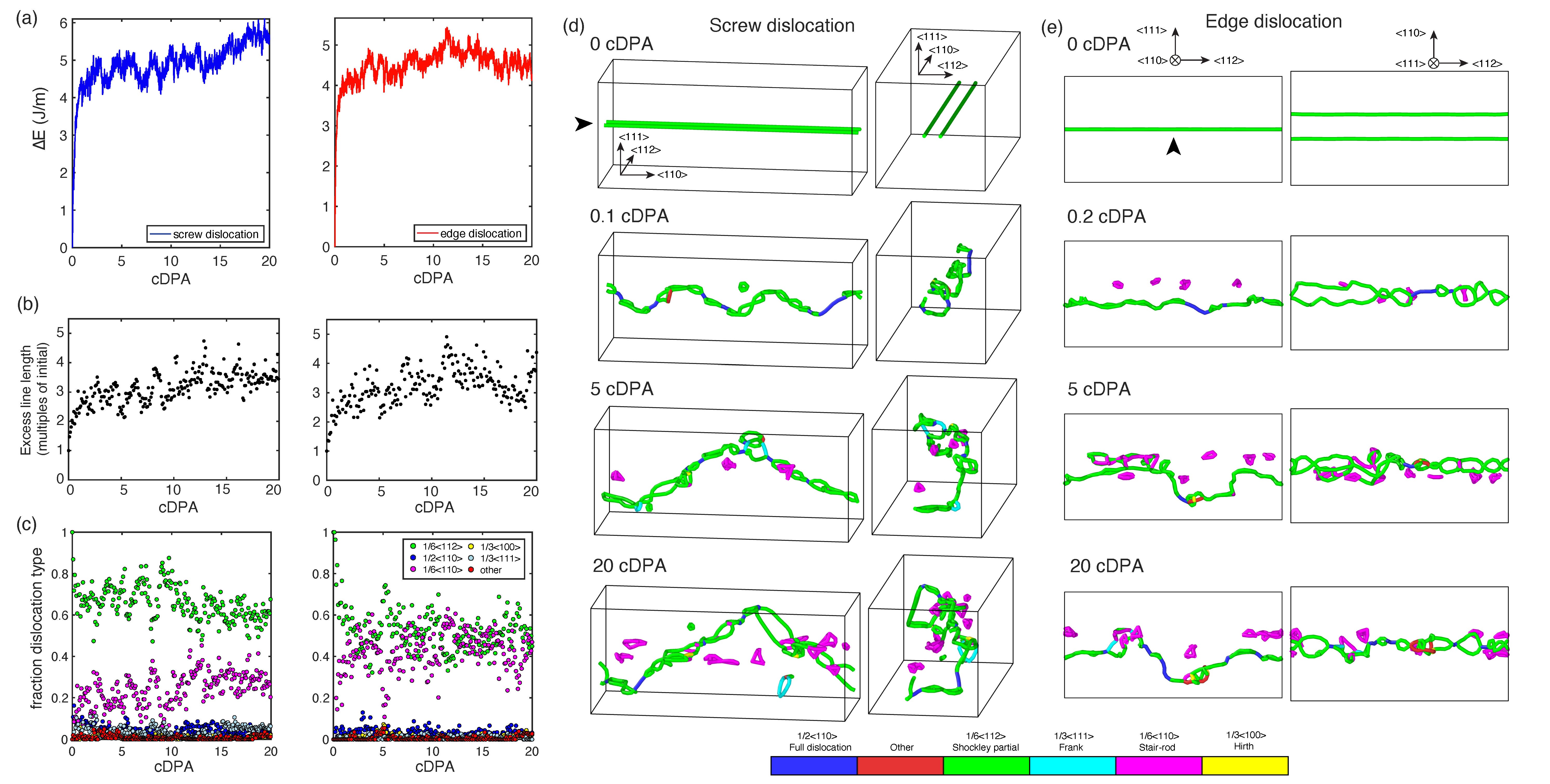}
\caption{CRA applied to screw and edge dislocations in FCC Ni. Measured quantities include the (a) excess line energy relative to the initial configuration, the (b) excess dislocation line length in multiples of initial line length and (c) the fraction of distinct dislocation types in the system. Panels on the left side of (a)-(c) correspond to the screw dislocation and panels on the right side to the edge dislocation. Microstructural evolution is shown for (d) screw dislocations and (e) edge dislocations at doses up to 20 cDPA. Two views are shown for each simulation frame with crystallographic sample axes labeled. Black arrows indicate the orientation of the second view with respect to the first.}
\label{fig:CRA_dislo}
\end{figure}

Dislocations exhibit steady state structures upon continuous creation of FPs that vary with dislocation character. The evolution of dislocation structure under irradiation for a screw and edge dislocation is shown in Fig. \ref{fig:CRA_dislo}. At small doses of 0.1 cDPA, an initially flat and fully dissociated screw dislocation assumes a helical morphology with partial dissociation (Fig. \ref{fig:CRA_dislo}d). The helix results from the formation of fluctuations with edge character along the dislocation line which subsequently climb via vacancy and interstitial mediated processes. The helix can be characterized by its average turning radius and average pitch, or distance between turns. An initial helix with three turns at 0.1 cDPA is observed to coarsen in both radius and pitch so that, by 5 cDPA, a helix with a single turn is observed. In the single turn system, the pitch is the length of the simulation box in the periodic line direction (20 nm) and the diameter of the helix spans the dynamic cylindrical region of the simulation (8 nm). We note that the helix outgrows the 2 nm wide radiation zone which is held constant throughout the simulation and was centered on the initial straight screw dislocation. The formation and coarsening of the helix involve a substantial increase in excess energy and a 3-4 increase in excess dislocation line length in the system as shown in Fig. \ref{fig:CRA_dislo}a and \ref{fig:CRA_dislo}b (lefthand panels).  

We consider the dose range 2-15 cDPA corresponding to the plateau in excess energy in Fig. \ref{fig:CRA_dislo}a (left) to correspond to local steady state dislocation structure under the constraints of our finite radiation zone and system size. In this regime, the single turn helix maintains its structure via a combination of glide and climb mechanisms and via reactions with dislocation loops which are observed to nucleate both homogeneously within the bulk crystal and heterogeneously at the helix itself. At higher doses in the range 15-20 cDPA, a continuous increase in excess energy suggests a departure from the initial local steady state. We hypothesize that this behavior originates from the inability of the helix to expand further beyond the rigid boundary conditions to keep up with bulk dislocation loop production. An increase in bulk dislocation concentration in the dose range 15-20 cDPA is indicated by an increase in the line fraction of stair-rod dislocations with 1/6\hkl<110> character which are observed as segments of stacking fault tetrahedra (Fig. \ref{fig:CRA_dislo}c). Helical turns of the dislocation become more rectangular in the dose range 15-20 cDPA as the dislocation attempts to absorb additional bulk dislocations. 

The edge dislocation exhibits climb behavior under irradiation with a CRA steady state corresponding to a jogged dislocation (Fig. \ref{fig:CRA_dislo}e). At a small dose of 0.2 cDPA, the edge dislocation is observed to climb to the edge of the radiation zone. Meanwhile, stacking fault tetrahedra with stair-rod character forms in the bulk irradiated region. Segments of the dislocation climb back toward the bulk defects and react with them, resulting in a jogged dislocation line with unit jogs coalescing into superjogs. In comparison to the irradiated screw dislocation, a similar excess dislocation length is added at and near the edge dislocation between 3-4 times the initial dislocation length (Fig. \ref{fig:CRA_dislo}b, right). A higher proportion of stacking fault tetrahedra are present at a steady state surrounding the edge dislocation compared to the screw dislocation (Fig. \ref{fig:CRA_dislo}c), suggesting less increase in length of the edge dislocation itself under irradiation compared to the helical screw dislocation. Local maxima in excess energy between 2-20 cDPA in Fig. \ref{fig:CRA_dislo}a correspond to maxima in the stair-rod dislocation concentration in Fig. \ref{fig:CRA_dislo}c. The peaks and valleys in these figures capture the cycle of bulk dislocation production and reaction with the mobile edge dislocation at a steady state. We emphasize that the aforementioned dislocation climb motion is driven purely athermally by the developing internal stresses due to the CRA protocol, as will be any structural change observed in the present work.

\subsection{Irradiation response of low angle grain boundaries}

\begin{figure}[ht!]
\centering\leavevmode \includegraphics[width=0.99\textwidth]{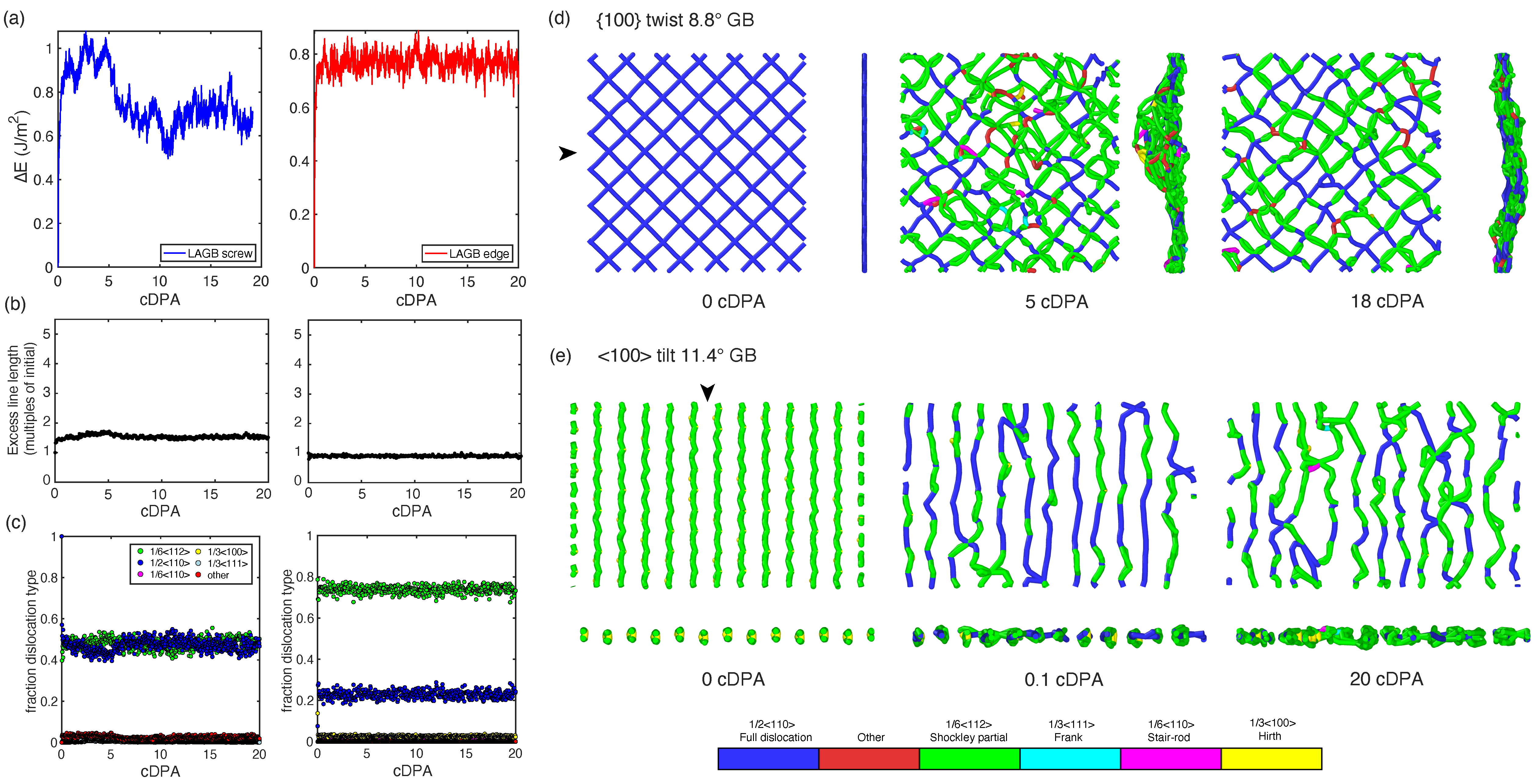}
\caption{CRA applied to selected low angle GBs in FCC Ni with tilt (edge) / twist (screw) character. Measured quantities include (a) excess interface energy relative to the initial configuration, (b) excess dislocation line length in multiples of initial line length and (c) the fraction of distinct dislocation types within the system. Panels on the left side of (a)-(c) correspond to the twist GB and panels on the right side to the tilt GB. Microstructural evolution is shown for (d) the twist GB with a \hkl{100} 8.8$^\circ$ misorientation and (e) the tilt GB with a \hkl<100> 11.4$^\circ$ misorientation. Two views are shown for each simulation frame. Black arrows indicate the orientation of the second view with respect to the first. }
\label{fig:CRA_LAGB}
\end{figure}

Low angle GBs (LAGBs) exhibit well-defined dislocation networks which accommodate a misorientation between two grains. The radiation response of LAGBs is expected to be intermediate between the behavior of single dislocations and high angle GBs. Building upon the application of CRA to single screw and edge dislocations, we consider the irradiation of screw and edge dislocation networks comprising low angle symmetric twist and tilt GBs. Fig. \ref{fig:CRA_LAGB} summarizes the irradiation responses of a $\hkl{100}$ twist 8.8$^\circ$ GB and a $\hkl<100>$ tilt 11.4$^\circ$ GB (ID 251 and 330 in \cite{olmsted2009survey}) at doses up to 20 cDPA. A network of perfect screw dislocations is present in the pristine twist GB, while the pristine tilt GB comprises a network of perfect edge dislocations dissociated into Shockley partial dislocations and Hirth locks (see Fig. \ref{fig:CRA_LAGB}d and e). 

The irradiated dislocation networks exhibit similarities and differences compared to irradiated single dislocations. The magnitude of dislocation lengthening is reduced in the case of LAGBs compared to single dislocations (Fig. \ref{fig:CRA_LAGB}b). The irradiated twist GB has a modest excess line length of 1.5 at 20 cDPA while the tilt GB exhibits no excess line length. In the twist GB, the excess line length is accommodated by the dissociation of full dislocations and by the out of plane glide and climb motion of the dislocation network. Individual dislocation segments bow out between junctions but do not form helices. In Fig. \ref{fig:CRA_LAGB}d, a bulge in the twist GB formed by local migration of the extended dislocation network is found to correspond to a local maximum in excess line length and excess energy at around 5 cDPA. The bulge is observed to move along the plane of the GB and eventually flatten via a series of dislocation reactions, leading to a drop in excess energy. Some residual curvature is left at the GB that persists in a fluctuating manner up to doses of 20 cDPA. This example shows how local GB migration can delay the approach to a steady state or can induce sudden transitions between two apparent steady state regimes of excess energy. In the tilt GB, climb-like motion of edge dislocations is observed in the GB plane, leading to the lengthening of individual dislocations, but this effect is offset by the reaction of relatively closely spaced dislocations and the recombination of partial dislocations, leading to no increase or even a slight decrease in total dislocation length. No out-of-the-plane motion of the tilt GB is observed. Overall, the response of LAGBs to irradiation is found to be more constrained than the response of single dislocations with a lesser magnitude of dislocation roughening and migration. Like the single dislocation case, the dislocation network character impacts possible climb mechanisms, and it is observed that twist GBs are likely to exhibit out-of-plane roughening events while tilt GBs exhibit in-plane roughening of GB dislocations.  

\subsection{Irradiation response of high angle grain boundaries}

High angle GBs (HAGBs) with misorientation angles above around $15^\circ$ have closely spaced GB dislocations in the GB core region, which are difficult to segment using DXA\cite{DXA}, especially when additional disorder is present due to radiation damage. We pursue the analysis of irradiated HAGB structures via distributions of interstitial atoms identified by Wigner-Seitz analysis. This analysis is most effective for GBs that do not migrate under irradiation since crystal reorientation leaves a signature in the interstitial distribution, which is difficult to decouple from other radiation damage effects. 

\begin{figure}[ht!]
\centering\leavevmode \includegraphics[width=0.99\textwidth]{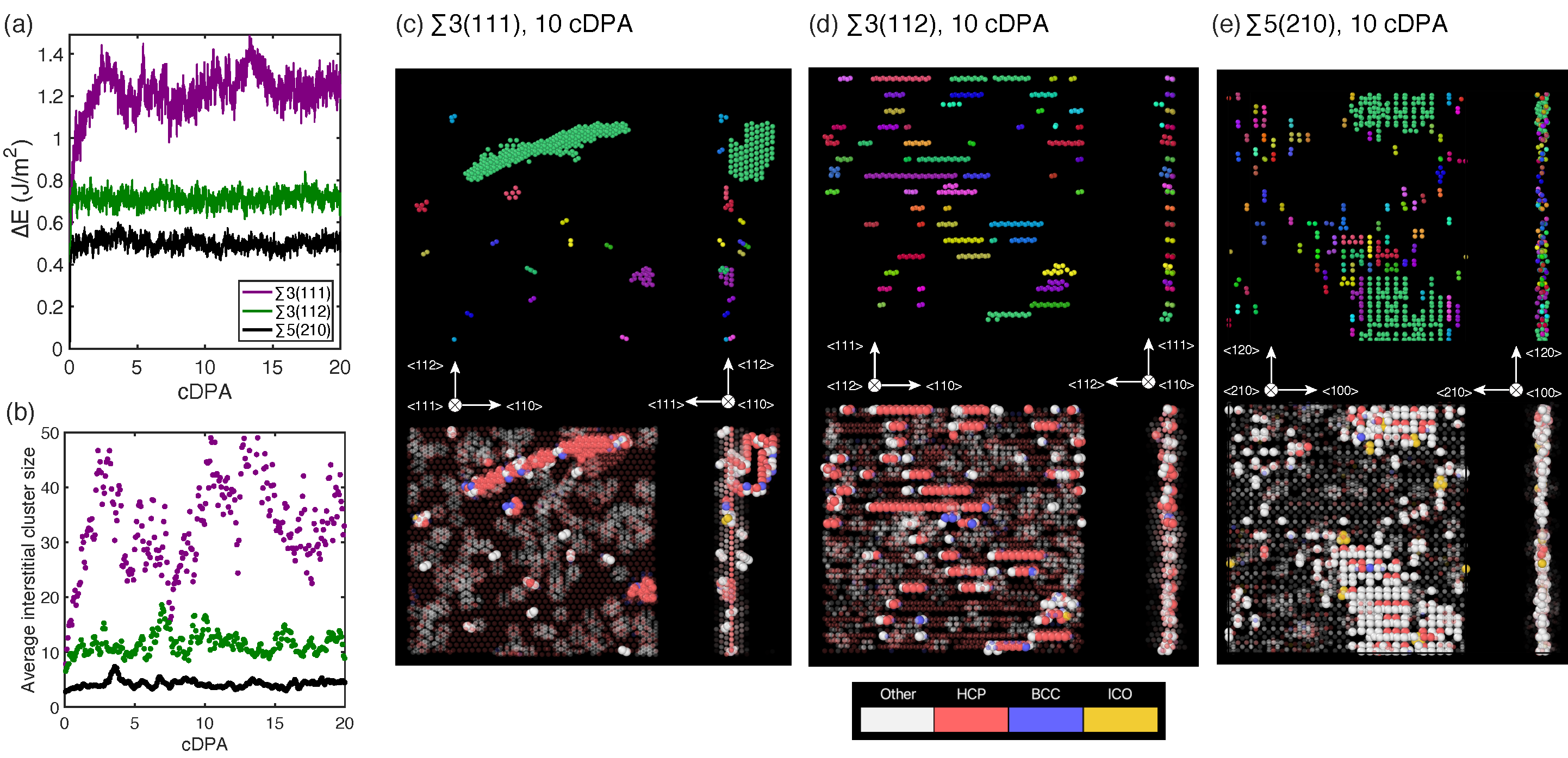}
\caption{CRA applied to selected high angle symmetric tilt GBs in FCC Ni. (a)-(b) show the change in excess energy and the variation of average interstitial cluster with dose. Interstitial clusters from Wigner-Seitz analysis are shown in (c)-(e) colored by cluster (upper panels) or local structural environment (lower panels). In the lower panels, transparent atoms are non-FCC atoms which are not interstitials, many of which are vacancies.}
\label{fig:CRA_HAGB_tilt}
\end{figure}

Fig. \ref{fig:CRA_HAGB_tilt} illustrates interstitial distributions under irradiation for several high angle tilt GBs with varying GB character. Interstitials created near GBs show a strong tendency to segregate to the GB core and form clusters. Vacancies exhibit weaker segregation behavior with supersaturation of vacancies remaining in the bulk crystal layers adjoining the GB. Interstitial cluster shape and size distributions at steady state depend on GB character. The evolution of average cluster size with irradiation is shown in Fig. \ref{fig:CRA_HAGB_tilt}b. For the three GBs shown, the average cluster size is inversely correlated with GB energy. Local maxima in interstitial cluster size with respect to dose correspond to local maxima in excess energy under irradiation (Fig. \ref{fig:CRA_HAGB_tilt}a). While the low energy $\Sigma 3 \hkl(111)$ coherent twin GB exhibits the formation of an interstitial platelet protruding from the GB which can be identified with a dislocation loop of predominantly Shockley partial character, the two other HAGBs exhibit interstitial clusters in the core of the GB which are more difficult to interpret in terms of dislocations. We note that interstitials can assemble into superstructures in the GB plane such as distinct GB phases \cite{frolov2013structural,hatton2024he} and disconnections, but that interstitial clusters can also form during processes such as the formation and motion of jogs at GB dislocations. In the case of the $\Sigma 3 \hkl(112)$ incoherent twin GB, interstitial clusters are elongated along the $\hkl<110>$ tilt axis. The $\Sigma 5 \hkl(210)$ GB, on the other hand, shows a single large cluster with less obvious shape anisotropy with respect to the tilt axis. No step is observed at the interface, ruling out the formation of a disconnection. A general picture which emerges is that irradiated GBs contain interstitial-rich and vacancy-rich regions which fluctuate in position at a steady state to accommodate continued irradiation. These point defect distributions are expected to be the result of core-level GB dislocation climb processes which couple to collective migration and reactions with bulk defect content. More sophisticated tools are needed to track the detailed core evolution of HAGBs under irradiation. We find that atomic coordination, shown for interstitial clusters in Fig. \ref{fig:CRA_HAGB_tilt}, does not distinguish interstitial rich regions from vacancy rich regions in a visually meaningful way, with the possible exception of icosahedral coordination occurring preferentially at the edges of interstitial clusters in the $\Sigma 5 \hkl(210)$ GB. 

\begin{figure}[ht!]
\centering\leavevmode \includegraphics[width=0.99\textwidth]{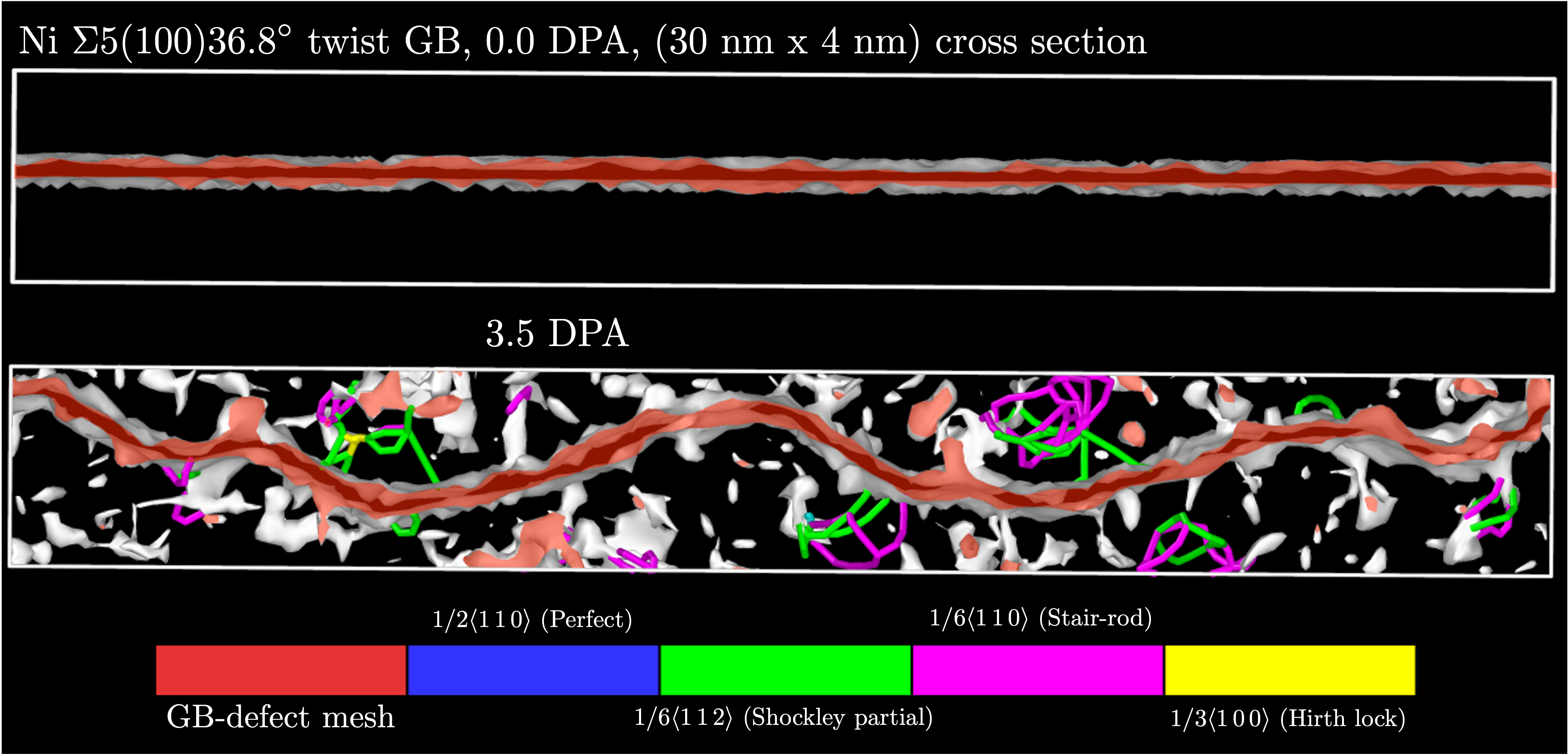}
\caption{CRA applied to a high angle twist GB in FCC Ni. Roughening of the GB plane is apparent.}
\label{fig:CRA_HAGB_twist}
\end{figure}

Similar to the analysis of LAGBs with twist character, we find that irradiated high angle twist GBs can exhibit significant changes in morphology and interface area under irradiation. One example is given in Fig. \ref{fig:CRA_HAGB_twist} for a $\Sigma 5 \hkl{100}$ twist GB with a relatively large GB plane size. The steady state GB morphology at 3.5 cDPA is observed to be wavy, indicating roughening of the initially flat GB. In this case, an irradiation zone size of 4 nm spanning the GB was chosen to allow for enhanced GB migration and roughening. A more detailed study of GB roughening under irradiation is left to future work. 

\begin{figure}[ht!]
\centering\leavevmode \includegraphics[width=0.99\textwidth]{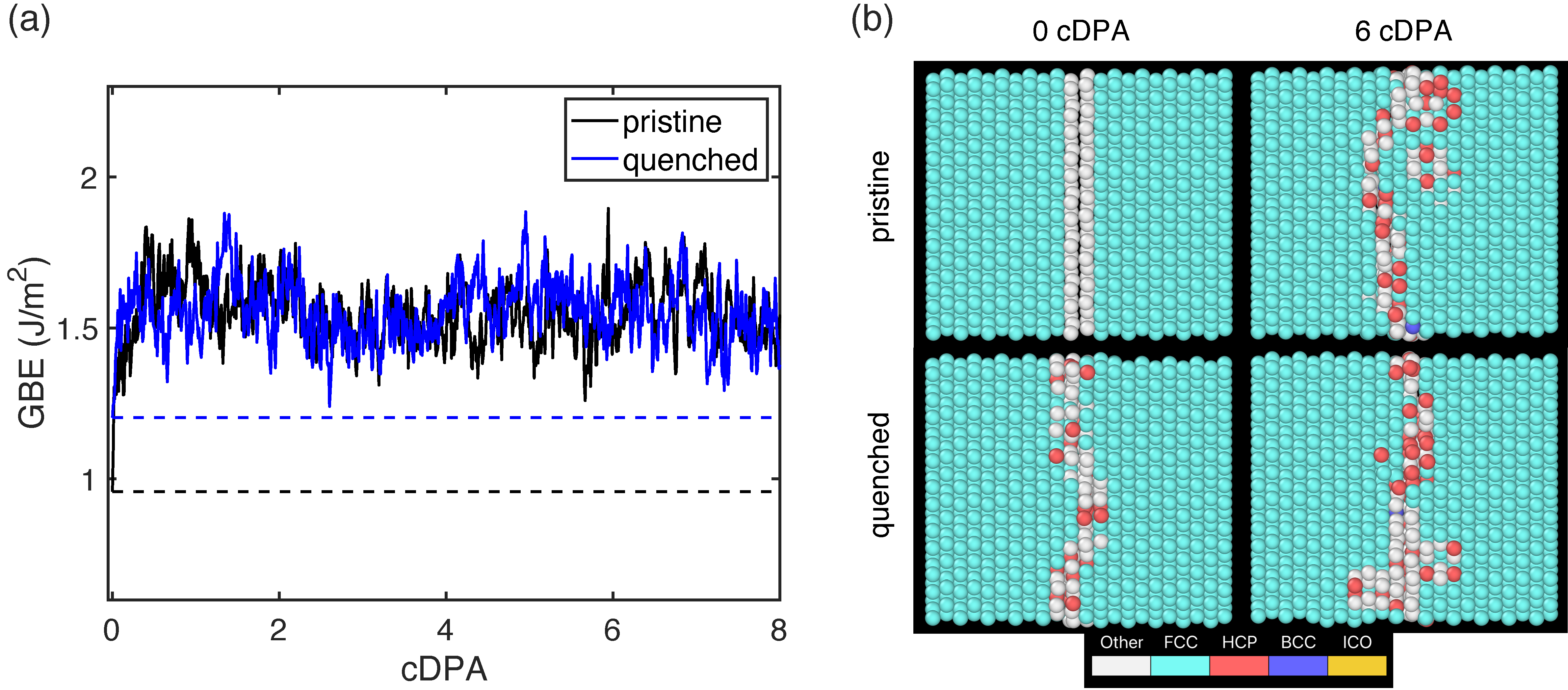}
\caption{CRA as an attractor for metastable GB structures. (a) Initially different structures (with initial GB energies given by dashed lines) follow statistically similar CRA trajectories. (b) Irradiated structures are visually similar for initially minimum energy and metastable structures.}
\label{fig:meta2}
\end{figure}

The CRA steady state is found to be an attractor for metastable variants of the same macroscopically defined GB with similar atomic densities. Fig. \ref{fig:meta2} illustrates that a minimum energy twist GB structure and a high energy metastable structure quenched from a glassy state reach statistically similar steady state GB structures under CRA in terms of saturation of excess energy and local atomic coordination. Similar results were obtained for the $\Sigma 5 \hkl(210)$ and $\Sigma 3 \hkl(112)$ tilt GB (not shown) to check the generality of the effect. During the creation of the high energy quenched structure, the target density of atoms is set to be equivalent to the average density of the pristine GB structure in the deposition zone. If the density is changed significantly, the resultant GB structure will be altered, and the steady state structure will be modified. In other words, the CRA steady state is an attractor for metastable GB structures with the same average density in the GB core region but is expected to vary for GBs with significantly different average densities in the GB core region. We hypothesize that there is a small range of GB core densities over which the same CRA steady state remains an attractor because of local swelling normal to the GB which can mediate volume and density changes approaching the steady state.  

\begin{figure}[ht!]
\centering\leavevmode \includegraphics[width=0.99\textwidth]{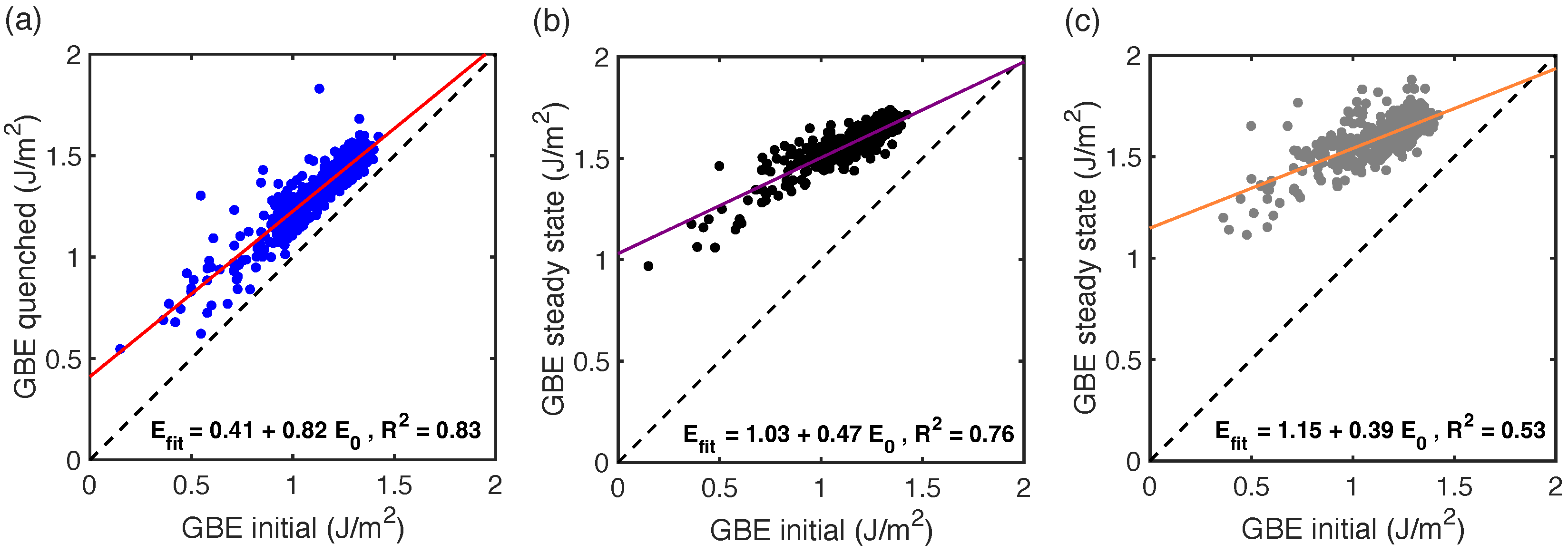}
\caption{Correlation between initial GB energies and perturbed GB energies. Perturbations are (a) quench from a glassy structure, (b) CRA applied to small systems: (3 nm x 3 nm) GB plane size, and (c) CRA applied to larger systems: (10 nm x 10 nm) GB plane size. Best fit lines and correlation coefficients are shown. Energies from the CRA simulations are computed at steady state with exact ranges which vary depending on the GB but typically span 3-10 cDPA.}
\label{fig:GBE}
\end{figure}

Fig. \ref{fig:GBE}b-c demonstrate a positive correlation between GB energies for pristine GBs and irradiated GBs at steady state. For comparison, Fig. \ref{fig:GBE}a shows the energies of GB structures quenched from a glass-like structure. A central result of this work is that high energy GBs reach a steady state with a smaller change in energy than low energy GBs. Radiation damage skews the GB energy landscape by raising the energy of low energy GBs relative to high energy GBs. This skewing effect is stronger for irradiated GBs than quenched GBs. Irradiated GBs also exhibit higher overall energies than quenched GBs. An explanation for these effects is that FP loading enhances defect formation in GBs relative to quenched structures by driving the separation of interstitial and vacancy populations and increasing overall vacancy and interstitial concentrations. The irradiated GBs maintain a larger non-equilibrium concentration of vacancies in the bulk crystalline regions surrounding the GB than the quenched GBs. 

The correlation between pristine and irradiated GB energy is slightly weaker for the large versus small GBs with different plane sizes but equivalent radiation zone widths. On average, steady state GB energy increases for the larger systems compared to the smaller systems. Size effects are system dependent but can involve interactions of defects such as disconnections or absorbed dislocations across periodic boundary conditions. In general, larger GBs exhibit higher densities of bulk dislocations in the adjoining crystals and more significant reconstructions, such as roughening, compared to small GBs. For large GBs, the true excess core energy of the GB is difficult to decouple from the excess energies of bulk dislocations immediately surrounding and attached to the GB. We view the small systems as the limiting case of homogeneous GB core reconstruction under irradiation at the level of one or several coincident site lattice unit cells with minimal generation of bulk defects. 

\subsection{Migration of defects after irradiation}

To survey the extent to which irradiation impacts materials properties, critical yield stresses and flow stresses were computed for strain driven migration of pristine and irradiated dislocation and GB structures. A dose of 5 cDPA was chosen for irradiated dislocation structures and 10 cDPA for GB structures. These doses were within the steady state damage regimes for the respective defect structures. 

\begin{figure}[ht!]
\centering\leavevmode \includegraphics[width=0.99\textwidth]{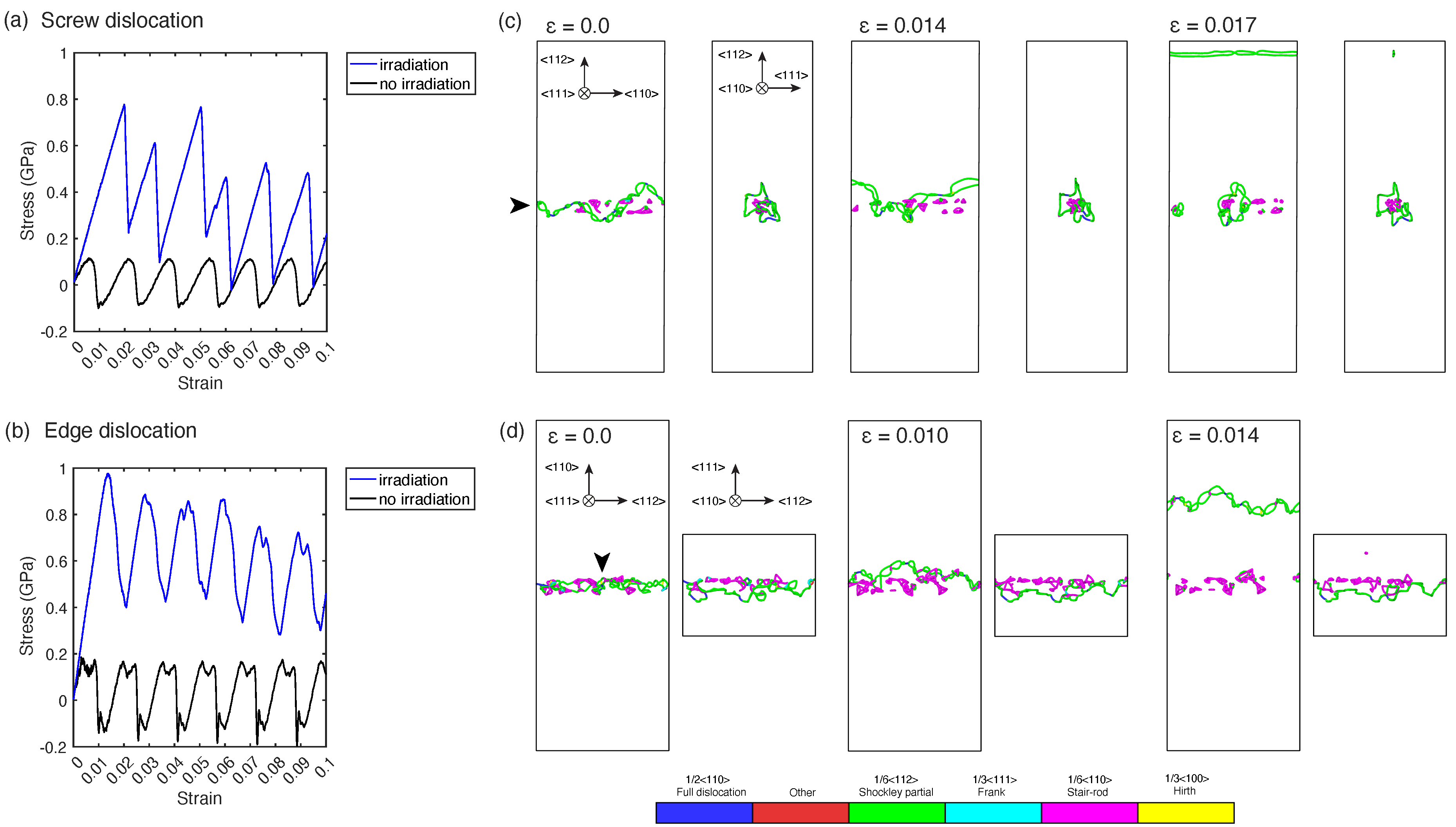}
\caption{Migration of pristine and irradiated dislocations. (a) and (b) show stress-strain curves for dislocation migration under shear loading. (c) and (d) show the initial yield/breakaway event with snapshots taken on either side of the first stress drop. Two views are shown for each simulation frame. Black arrows indicate the orientation of the second view with respect to the first.}
\label{fig:DM}
\end{figure}

Irradiation-induced damage in the vicinity of dislocations is found to pin the motion of dislocations, raising the critical yield stress for dislocation motion relative to pristine dislocations. Fig. \ref{fig:DM} illustrates the migration response of pristine and irradiated screw and edge dislocations with stress-strain curves given in Fig. \ref{fig:DM}a-b. Stress drops in the mechanical response of irradiated dislocations correspond to de-pinning events where dislocations break away from their radiation zones. Between each stress drop, dislocations migrate through the bulk crystal across periodic boundaries and become pinned within the radiation zones again. The magnitude of irradiation-induced hardening is found to be slightly larger for the edge dislocation than screw dislocation, as can be seen from the differences in peak stresses of the irradiated versus pristine samples, around 0.8 GPa for the first stress peak of the edge dislocation and 0.7 GPa for the first stress peak of the screw dislocation. De-pinning dynamics differ between the screw and edge dislocation. The helical screw dislocation migrates by shedding a large dislocation loop and recovering its initial pristine structure with nearly straight dislocation lines apart from a few constrictions. The edge dislocation maintains a jogged structure upon breakaway from the radiation zone. We do not have evidence for a change in mobility upon de-pinning in the irradiated versus pristine dislocations. Small serrations in the stress strain curve corresponding to dislocation motion between Peierls valleys are practically identical in the pristine and irradiated samples, implying equivalently large mobilities in both cases. We note that large period serrations in the stress strain curves of the migrating pristine dislocations are an artifact of the fixed, displacement controlled boundary conditions of the simulation. The elastic strain in the system is incompatible with the plastic strain due to dislocation motion, resulting in an additional force on the dislocation. This issue is discussed further in \cite{rodney2007activation} and can be corrected via the use of flexible boundary conditions. The main conclusion regarding the larger flow stress of migration of the irradiated versus pristine dislocations is not impacted by this issue. 

\begin{figure}[ht!]
\centering\leavevmode \includegraphics[width=0.99\textwidth]{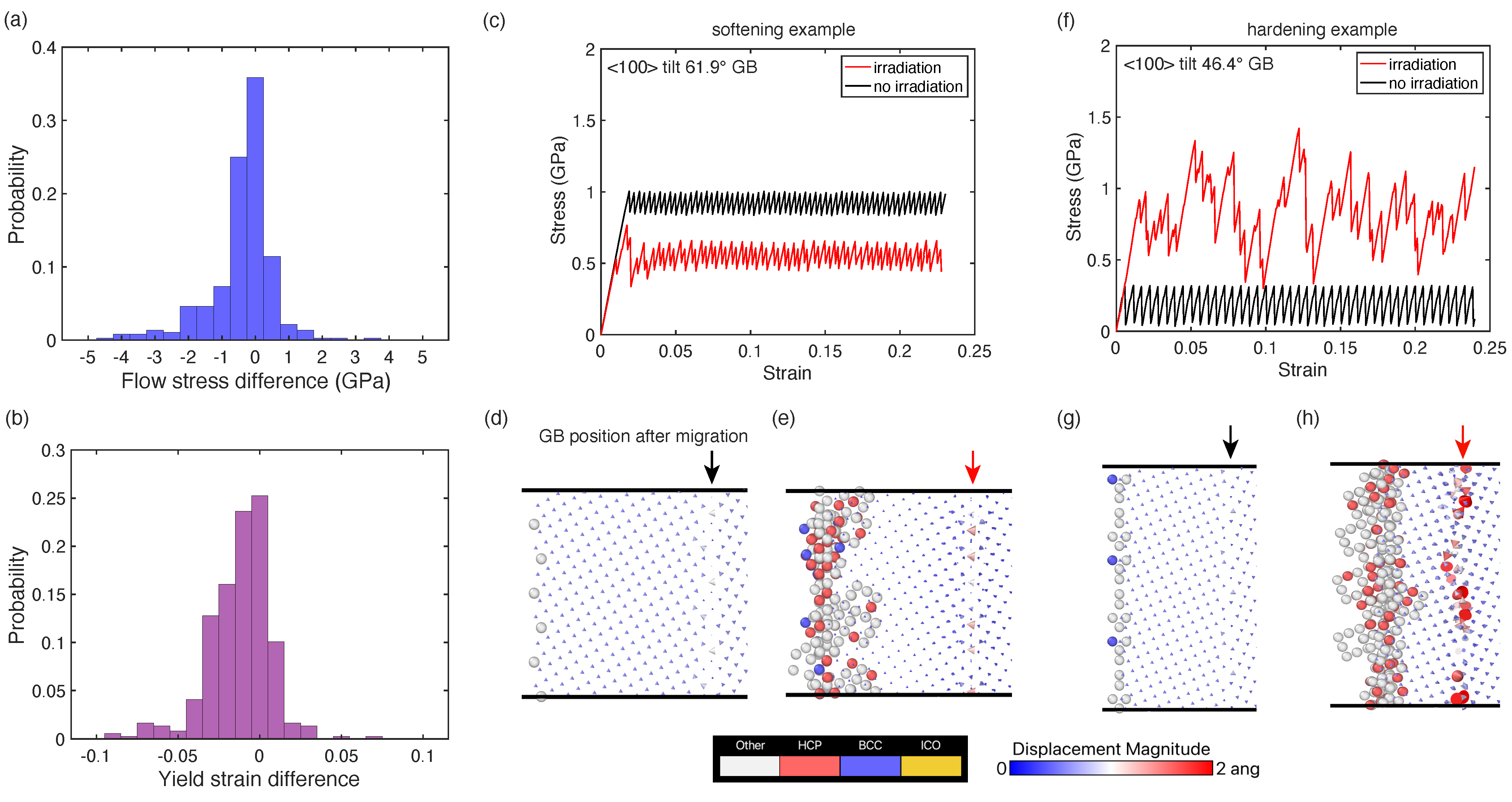}
\caption{Migration of pristine and irradiated GBs. (a)-(b) show statistics for all 388 Ni GBs. GBs exhibit a range of softening / hardening responses in flow stress (taken as average stress $>$ 0.15) depending on GB type. Examples of softening and hardening are shown in (c)-(e) and (f)-(h). In (d)-(e) and (g)-(h), images show the displacement field of a migrating GB over a small strain interval (corresponding to approximately one stress drop) superimposed on the structure of the GB (pristine or irradiated) before the application of the shear strain. Differences in displacement fields between the pristine and irradiated GBs indicate sustained differences in core structure which contribute to softening or hardening effects.}
\label{fig:GBM}
\end{figure}

In contrast to dislocations, irradiated GBs in Ni are found to have changes in mobility that are sustained upon leaving the radiation zone. While dislocations show a hardening effect for yield under irradiation, GBs can either harden or soften under irradiation depending on GB character. Serrations in the stress-strain curves for GB migration in Fig. \ref{fig:GBM}c,f have a different meaning than in Fig. \ref{fig:DM} for dislocations. Stress drops correspond to sequential migration events out of the radiation zone and the GB only breaks away from the radiation zone once. Average flow stresses are computed beyond shear strains of 0.15 at which point mobile GBs have left the radiation zone. It is worth noting that some GBs, such as pure twist GBs, exhibit pure sliding and never leave the radiation zone, though combinations of pure sliding and normal migration are possible. Fig. \ref{fig:GBM}a-b gives distributions of flow stress and yield strain changes for irradiated GBs compared to their plastically deformed pristine counterparts. The most common behavior for around 48\% of irradiated GBs is a softening effect in which yield strain and flow stress decrease compared to the pristine GBs. Around 36\% of GBs exhibit no or minimal change in flow stress, implying that they recover their initial pristine structure during migration. The final 16\% of GBs exhibit hardening behavior in flow stress as compared to the pristine GBs. 

A different flow stress for a mobile irradiated GB versus pristine GB implies an altered mobile core structure that is sustained outside of the radiation zone. Examples are given for two \hkl<100> tilt GBs in Fig. \ref{fig:GBM}c-f which show softening and hardening behavior during migration. In Fig. \ref{fig:GBM}d-e,g-h, the initial pristine and irradiated non-FCC GB core structures are superimposed on the displacement field of the migrating GBs over a strain interval of one stress drop. Different displacement fields distinguish the migration mechanisms of pristine and irradiated GBs, implying a sustained difference in core structure upon migration outside of the radiation zone. In the hardening case, the irradiated GB exhibits a rough morphology as it migrates with relatively long string-like displacements on the order of a nearest neighbor distance which are not observed in the migrating pristine GB. In the softening case, a change in core structure makes migration easier. The irradiated GB exhibits shuffle displacement vectors that are each a fraction of the nearest neighbor distance and have a different geometric character than the mobile pristine GB. We note that GBs can leave vacancies formed in the radiation zone behind during migration, moving forward with an altered density. This is one mechanism to sustain a different mobile core structure, though significant differences in flow stresses are still observed in cases where no vacancies are left behind such as the migration of the tilt GB in Fig. \ref{fig:GBM}c-e.

\section{Discussion}
\label{discussion}

This work sheds new light on the nature of GBs and dislocations as sinks and storage sites for radiation-induced point defects by proving the existence of irradiation-induced steady state defect structures in the athermal limit.  A main result of this work is that, on average and under identical irradiation conditions, low energy GBs have a larger change in energy required to reach a steady state than high energy GBs. This result agrees with the findings of Foley et al. from athermal simulations of continuous vacancy loading at symmetric tilt GBs \cite{NEGB_sim2}, though the applied doses in this work were larger and better defined because of mass conservation. The change in energy to reach a steady state can be understood as the maximum energy available to drive source/sink action at defects, or critical driving energy, a quantity that is known to impact GB sink efficiency \cite{sutton1995interfaces}. Balluffi et. al pointed out that different critical driving energies are expected to be required for different GB types to overcome restrictions to dislocation climb or add dislocation length \cite{sutton1995interfaces}. The critical driving energy is different from sink efficiency, a finite temperature quantity that is the solution to a diffusion equation with particular boundary conditions \cite{el2018does}. Defects with large critical driving energies are expected to show thresholding effects of sink efficiency depending on the magnitude of the driving energy which may vary with irradiation conditions. 

To put our findings in perspective with prior literature, we compare the taxonomy of radiation induced steady state microstructures in this work to existing observations  of radiation damage near pre-existing defects from experiments and simulations. A theme that emerges is that the athermal climb-induced microstructures observed in this work are metastable equilibria, susceptible to thermally activated recovery via atomic diffusion. Nevertheless, irradiation induced steady state local microstructures observed in this work are found to agree qualitatively with observed behavior in experiments in-situ during active irradiation or during frozen-in stages of kinetic development.  

Helical dislocations have been observed in a variety of high-purity metals, alloys, and ceramic materials arising from the interaction of pre-existing screw dislocations with large fluxes of point defects \cite{amelinckx1957formation, weertman1957helical, caslavsky1971observation, veyssiere1972possible, anderson2017theory, haley2019helical, li2023formation}. In FCC materials, helical dislocations have been observed in high stacking fault metals such as high purity Al and Al-Cu alloys \cite{thomas1959helical, anderson2017theory} with partial dislocation dissociation along the length of the helix. The formation of helices under irradiation has also been observed in body-centered cubic (BCC) materials in neutron-irradiated FeCr alloys and ion-irradiated FeCrAl alloys \cite{haley2019helical,li2023formation}. The free energy of formation of a helical dislocation involves a balance between energetic penalty terms such as dislocation line energy and the elastic interaction between coils and attractive chemical terms promoting dislocation climb-mediated absorption of point defects from a super-saturated solution \cite{anderson2017theory}. Discrete dislocation dynamic simulations have modeled the evolution of helical dislocations under a vacancy supersaturation through a combination of climb and glide mechanisms \cite{munday2016prismatic, liu2017numerical, erel2017generation}. Atomistic simulations have examined the unit mechanisms of absorption of prismatic dislocation loops at screw dislocations to form helical turns in FCC and BCC metals \cite{rodney2004molecular, robach2006dynamic, liu2008molecular} and have revealed the formation of helical dislocations via dislocation reactions in the vicinity of nano-precipitates \cite{cui2018new}. To our knowledge, our simulations are the first direct atomistic simulations of helical dislocation formation via continuous irradiation. The obtained helices have smaller radii and pitch than experimentally observed helices because of imposed boundary conditions and system size restrictions \cite{anderson2017theory}. Edge dislocations have been experimentally observed to climb and form superjogs in a variety of irradiated materials with prior atomistic and dislocation dynamics simulations capturing this behavior \cite{li2020evolution, rodney2000dislocation, clouet2006vacancy, terentyev2008simulation}. 

Experimental data on irradiation damage of dislocation networks comprising low angle GBs is scarce. Vicinal GBs with secondary GB dislocations superimposed on singular high angle GBs have been experimentally observed to undergo climb processes and absorb extrinsic dislocations under irradiation \cite{sutton1995interfaces, komem1972direct}. Models have been developed for radiation induced climb of GB dislocations in low angle tilt GBs \cite{sutton1995interfaces, gu2017point}. Less attention has been devoted to modeling the evolution of low angle twist GBs subject to irradiation. Szlufarska et. al simulated interstitial loading of low angle \hkl{100} and \hkl(111) twist GBs in SiC and observed dissociation and climb of intrinsic GB dislocations with dislocation reactions originating at dislocation junctions which act as traps for point defects \cite{jiang2017multiple}. Martinez et al. developed a hybrid kinetic Monte Carlo-MD algorithm to simulate the evolution of low angle \hkl(110) and \hkl(111) twist GBs in Fe and Cu under vacancy loading and found a change in dislocation structure of the Cu interface under vacancy aggregation \cite{martinez2012atomistic}. The details of the reactions observed in these prior works differ from the observed evolution of the \hkl{100} low angle twist GBs in this work and involve less significant roughening of the GB plane. An interesting observation in this work is that GB dislocations appear to favor dissociation in the twist GB and recombination in the tilt GB. This observation runs counter to intuition on the basis of repulsive interactions of partial dislocations. Partial dislocations comprising screw dislocations in FCC metals are known to have a weaker repulsive interaction compared to the partials comprising the edge dislocation \cite{anderson2017theory}, favoring constriction in the former case, especially for high stacking fault energy materials like Ni. Irradiation drives the system in the opposite direction in the LAGBs, possibly due to the formation of defective states that compensate for the elastic interactions. 

A number of experimental observations exist regarding the evolution of HAGB structures under irradiation  \cite{king1980mechanisms, dollar1985point, lee2021grain,wei2022direct, barr2022irradiation}. Nucleation and growth of dislocation loops at the coherent twin GB has been observed in many FCC metals either under irradiation or after quenching from high temperatures \cite{sutton1995interfaces,king1980mechanisms}. Barr et. al subjected a faceted incoherent twin GB in Pt to 1 DPA of ion beam damage and observed facet motion via presumed disconnection climb mechanisms \cite{barr2022irradiation}. Lee et. al subjected two GBs with the same misorientation in Au with near tilt and twist character to an unspecified dose of electron beam irradiation and observed faceting of the twist GB and roughening and migration of the tilt GB \cite{lee2021grain}. An interesting prediction from our work which awaits experimental validation is the roughening of GBs with \hkl{100}twist character under irradiation. 

To test the stability of steady state structures produced by CRA at finite temperature, selected samples, including the helical screw dislocation and several HAGBs with small system sizes, were annealed via MD simulations at high temperatures (0.8-0.9 times the melting point) where short-circuit diffusion mechanisms were observed to be active. At such high temperatures, initially, pristine and irradiated HAGB structures were found to be visually indistinguishable with nearly identical measured diffusivities, indicating recovery of irradiation damage. Such recovery processes are common in aging experiments in nano-crystalline materials, where aging of fs-pulse irradiated materials in the vicinity of GBs was found to lead to recovery of the strength of the un-irradiated material \cite{balbus2018femtosecond}. In contrast to the tested HAGBs, the helical dislocation was not observed to collapse or self-annihilate excess dislocation length within 40 ns of annealing, indicating structural stability on MD timescales even at high temperatures. The stability of the helical dislocation on MD timescales is likely a reflection of slow diffusion along the curved dislocation core relative to GB diffusion at these temperatures. The results of this work are expected to be most applicable to irradiation damage at low temperatures and high dose rates where stress-driven mechanisms dominate thermally activated mechanisms. 

The majority of GBs in this study exhibited softening or hardening effects during migration after irradiation, indicating that radiation damage can induce changes in defect core structure that are carried over long distances through a material. GBs act as long range vessels for point defect transport, but may also assist in defect annihilation, depending on GB character. Almost 30\% of GB structures in this study were found to recover their initial structures upon migration, similar to observations of GB migration under continuous vacancy or interstitial loading \cite{yu2015non, li2023vacancy}. Multiple studies have shown that migration mechanisms such as disconnection nucleation and growth can proceed with altered energy barriers or shear coupling factors in the presence of point defects \cite{combe2019heterogeneous, chen2020atomistic, anento2020interaction, chen2021associating}. Damage in the form of GB phase transitions can even change the activated disconnection mode entirely \cite{frolov2013structural, frolov2018structures}. The statistics in Fig. \ref{fig:GBM} represent a variety of plastic deformation mechanisms at GBs including coupled motion, sliding and emission of dislocations from GBs. The most common scenario for softening or hardening appears to be the preservation of the underlying migration mode in the pristine GB with altered energy barriers, as was observed in the tilt GB migration in Fig. \ref{fig:GBM}c-h. These results support the hypothesis that metastable structure variants of the same macroscopic GB with similar densities in the GB core region can exhibit different mobilities for the same mode of migration \cite{chesser2021optimal}.

\begin{figure}[ht!]
\centering\leavevmode \includegraphics[width=0.8\textwidth]{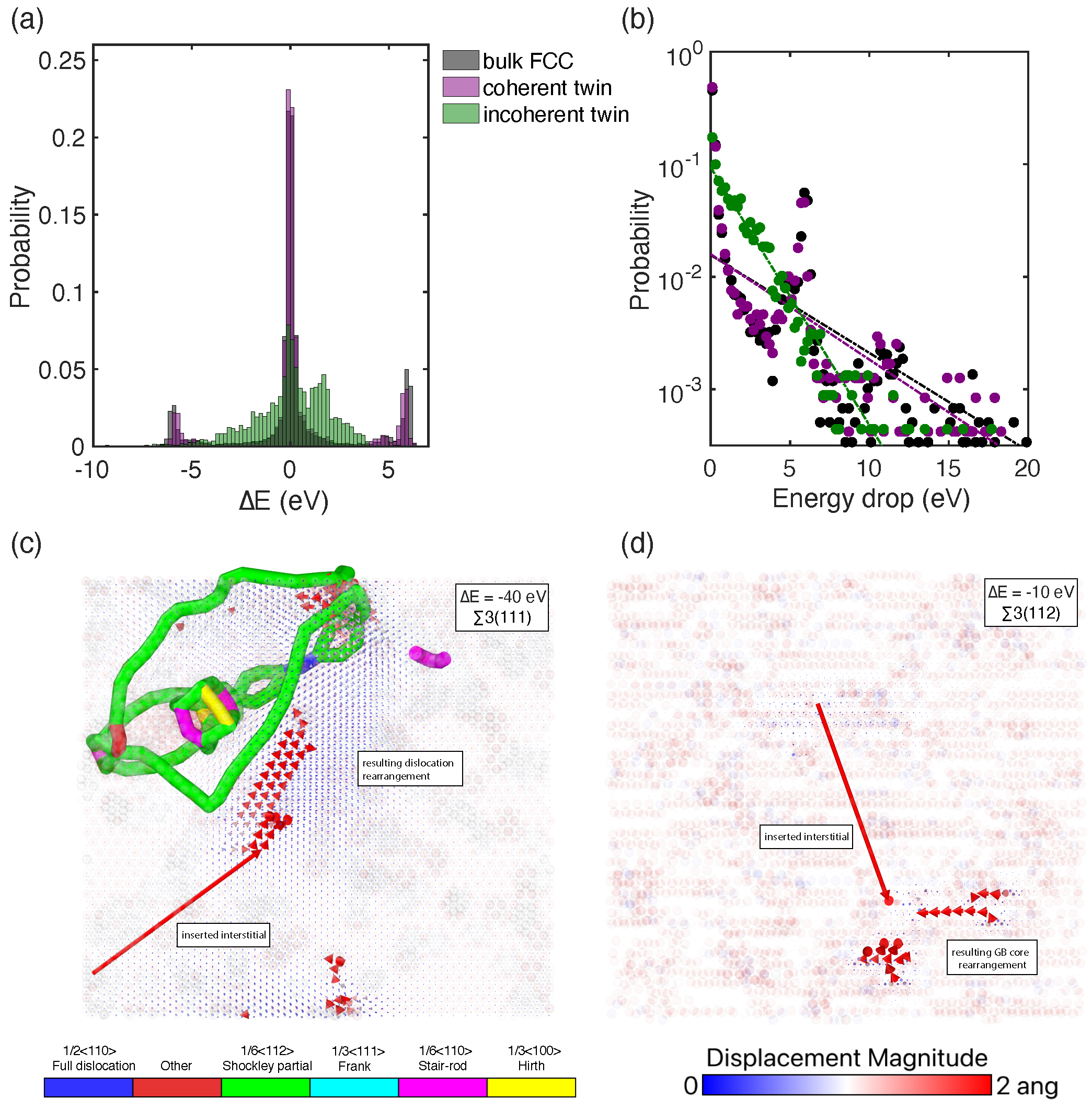}
\caption{Energy fluctuations at steady state. (a) Histogram of total energy changes for the insertion of 5000 single FPs at steady state for bulk FCC Ni and two GBs. (b) Energy drop distributions for each system with fits to exponential decay demonstrate bulk-like behavior for the coherent twin GB and glass-like behavior for the incoherent twin. (c)-(d) Displacement fields projected onto the GB plane for large example energy drop events at the coherent and incoherent twin GBs. The crystal frame directions are the same as in Fig. \ref{fig:CRA_HAGB_tilt}.}
\label{fig:exp}
\end{figure}

GBs are found to exhibit irradiation dynamics at a steady state intermediate between bulk crystals and glasses. GBs are similar to glasses in the sense that their disordered steady state structures under irradiation are attractors for initial metastable structures with varying degrees of relaxation. Prior CRA studies of bulk materials have analyzed relaxation statistics for individual FPs inserted into the irradiated material at steady state and have found non-Gaussian distributions of energy changes reflecting correlated and collective responses to FP insertion in crystals and glasses \cite{CRA_BCC, CRA_glass}. Bulk BCC Fe crystals were found to exhibit relaxation spectra with power law distributed energy drops \cite{CRA_BCC}, whereas an irradiated glass exhibited exponential decay in its energy drop spectrum \cite{CRA_glass}. A physical interpretation of these results is that power law decay corresponds to rare large-scale collective events such as dislocation network rearrangement, whereas exponential decay corresponds to more local events involving the collective displacement of fewer atoms. The relaxation spectra for selected GBs at steady state are shown in Fig. \ref{fig:exp} with comparison to relaxation spectra for bulk FCC Ni. Microstructural evolution for bulk FCC Ni under CRA is summarized in Supplementary Fig. S1. At steady state, bulk FCC Ni exhibits a dislocation network with a large fraction of stair-rod dislocations, mainly in the form of truncated stacking fault tetrahedra. Smaller fractions of Shockley partial dislocations and Frank loops are also present. Total energy changes at steady state in Fig. \ref{fig:exp} show well-defined peaks in the same positions for bulk FCC Ni and the coherent twin GB. Peaks at $\pm$5.9 eV correspond to the sum of individual interstitial and vacancy formation energies of 4.2 eV and 1.7 eV. Unlike the coherent twin, the incoherent twin GB has a wide distribution of energies and shows a peak at 1.7 eV rather than 5.9 eV corresponding to the bulk vacancy formation energy.  This peak at 1.7 eV does not have a symmetric equivalent on the negative side of the distribution, reflecting a wide range of possible core-level vacancy annihilation processes. The continuous distribution of total energy changes in the incoherent twin GB is reminiscent of the continuous distribution of point defect formation and migration energies previously computed for high energy pristine and non-equilibrium GBs \cite{bai2020mapping}. 

The energy drop spectrum for the incoherent twin GB is well fit by an exponential decay (\ref{fig:exp}b), indicating a glass-like relaxation behavior which is representative of the radiation response of high energy GBs. Exponential fits are also shown for the coherent twin GB and bulk FCC Ni, though the energy drop distributions for these systems are found to be a better fit to a power law decay (for sufficiently large bin sizes that capture peak behavior). Examples of large relaxation events in response to FP insertion are shown for the coherent twin GB and an incoherent twin GB in \ref{fig:exp}c-d. A 40 eV energy drop within the coherent twin GB corresponds to a collective dislocation migration and vacancy redistribution event, whereas a 10 eV energy drop in the incoherent twin corresponds to correlated atomic rearrangements in the vicinity of the inserted interstitial atom. We note that a rearrangement in one location can trigger nearby rearrangements a short distance away, illustrating the concept of dynamical facilitation \cite{chandler2010dynamics, chacko2021elastoplasticity}. String-like relaxation events in the incoherent twin GB resemble short-circuit diffusion mechanisms observed for this GB at finite temperature, though at intermediated homologous temperatures the length-scale of correlated avalanche-like diffusion events are larger than observed here \cite{chesser2022point, chesser2024atomic}. The existence of dynamical facilitation observed at 0 K in this work is helpful for isolating a poorly understood aspect of GB diffusion that is difficult to study at finite temperature \cite{chacko2021elastoplasticity}. Overall, GBs exhibit a wide range of collective stress-relaxation mechanisms at steady state, including local migration, dislocation reaction events, and string-like displacement events, with high energy GBs exhibiting a continuous spectrum of relaxation events well fit by an exponentially decaying distribution. 

\begin{figure}[ht!]
\centering\leavevmode \includegraphics[width=0.8\textwidth]{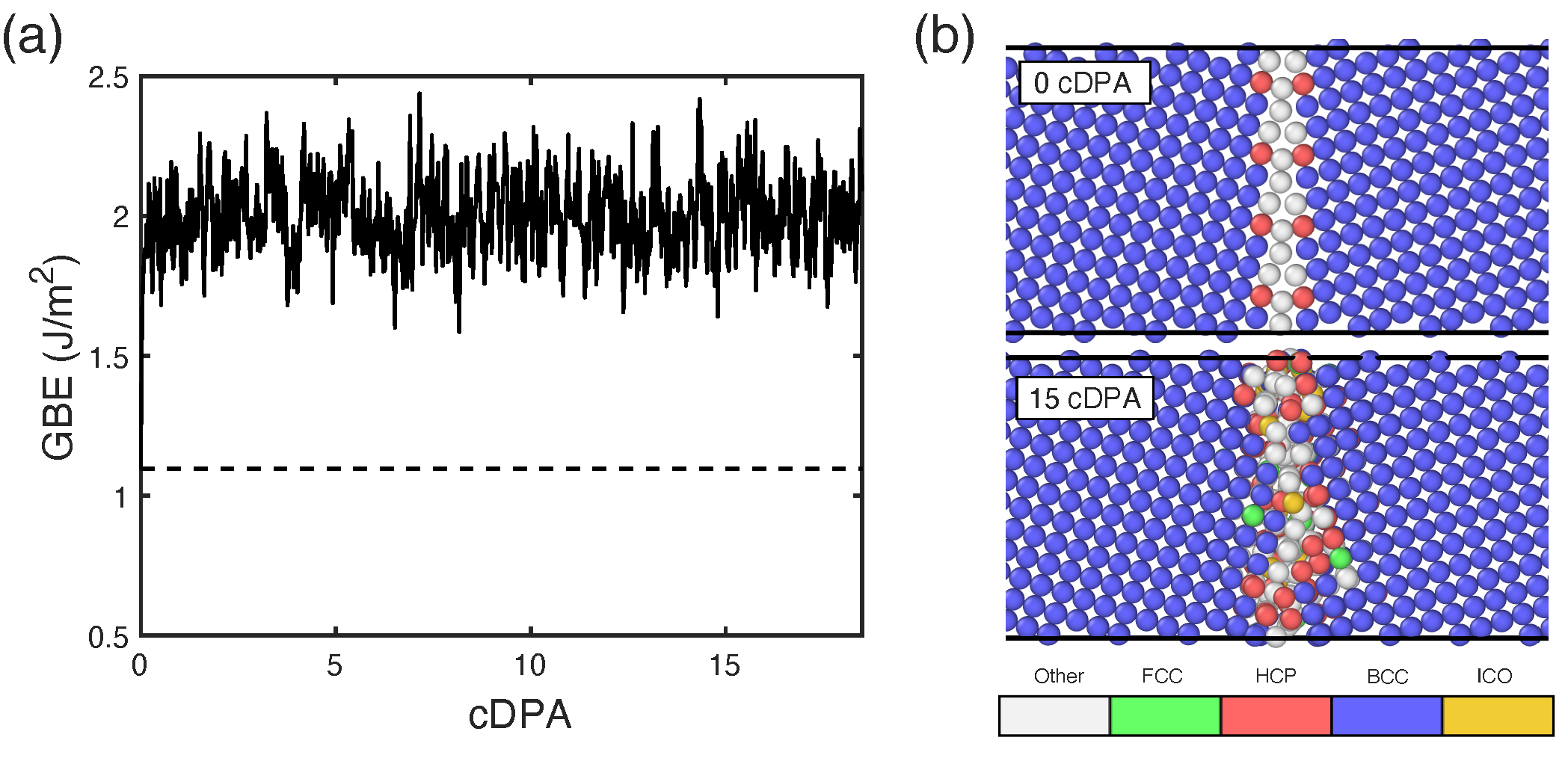}
\caption{CRA applied to the $\Sigma5\hkl(310)\hkl<100>$ tilt GB in BCC Fe. (a) Excess GB energy as a function of radiation dose with (b) images of the pristine GB core structure and irradiated system at 15 cDPA.}
\label{fig:Fe}
\end{figure}

\begin{figure}[ht!]
\centering\leavevmode \includegraphics[width=0.99\textwidth]{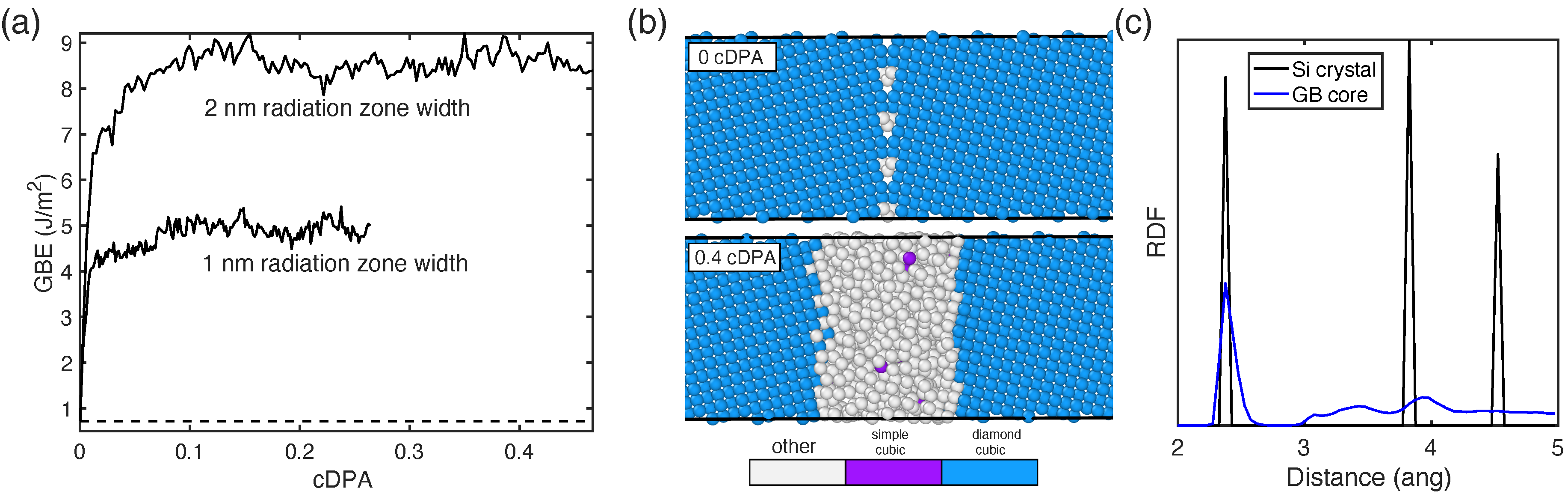}
\caption{CRA applied to the $\Sigma17\hkl(530)\hkl<100>$ tilt GB in diamond cubic Si. (a) Excess GB energy as a function of radiation dose for two zone widths of 1 nm and 2 nm. (b) Snapshots of the simulation with the larger zone size at 0 and 0.4 cDPA reveal continuous disorder in the irradiated region. (c) Radial distribution function (RDF) for a slab of Si crystal and the disordered GB core region under irradiation.}
\label{fig:Si}
\end{figure}

We expect CRA steady state microstructures in the vicinity of defects to be a generic feature of microstructural evolution in crystalline materials. However, the particular features of microstructural evolution observed in this work for FCC Ni depend on the disparity between relatively small bulk interstitial migration energies and relatively large vacancy migration energies in FCC Ni. In the case of Ni and other close-packed metals with low bulk interstitial migration energies, GBs are expected to act as strong sinks for interstitial atoms and avoid amorphization under irradiation (even though they can still exhibit glassy dynamics). An example of steady state irradiation behavior for a $\Sigma5\hkl(310)\hkl<100>$ symmetric tilt GB in Fe is shown in Fig. \ref{fig:Fe} using the Mendelev Fe potential \cite{mendelev2003development}.  An example of the irradiation response of a  $\Sigma17\hkl(530)\hkl<100>$ symmetric tilt GB in Si is also shown for comparison in Fig. \ref{fig:Si} using the Mishin Si Tersoff potential \cite{pun2017optimized}. In Fe, the GB structure reaches and maintains a steady-state disordered GB structure in a manner that is qualitatively similar to GBs in Ni. The GB core thickens slightly and takes on a rougher morphology, but does not amorphize. On the other hand, the irradiated Si GB structure forms a disordered structure spanning the irradiation zone with a direct scaling of interface energy with the width of the irradiation zone that is not present in the surveyed close-packed metals. The radial distribution function for atoms in the disordered core region suggests an amorphous structure. We note that amorphous intergranular films are known to have high radiation damage tolerance and that GB amorphization is one strategy to improve radiation damage tolerance\cite{schuler2020amorphous,aksoy2024enhanced}. 

This work raises fundamental questions that present opportunities for future studies. CRA is a computationally efficient method to investigate the saturation of topological defect core structures under irradiation in the athermal limit. More work is needed to understand the system size, radiation zone size, and material dependence of CRA in the presence of pre-existing defects. The size effect observed for CRA steady states with respect to GB plane size should be better understood. In the limit of large GB or irradiation zone sizes, does initial GB energy lose its predictive power for determining features of the steady state such as saturation energy? Can methods be developed to de-convolute the energy contributions of different defect types at steady state or more quantitatively analyze the changes in HAGB core structure? Low angle GBs present interesting opportunities to study dislocation network climb behavior in more detail. Do LAGBs with sufficiently small twist angles form helices between dislocation junctions under irradiation? In this work, we only scratched the surface of the GB character dependence on radiation damage. Future work can examine trends in radiation damage with boundary plane character and study irradiation-induced migration in more depth. 

There are further opportunities to assess changes in materials properties due to irradiation-induced GB metastability \cite{el2024grain}. In the Supplementary Information, we present results of nanopillar compression tests \cite{mishra2023role} for two pristine and irradiated bicrystal structures in Ni and Fe (see Supplementary Fig. S2 and S3). The impact of GB metastability on the yield and flow stresses of the bicrystals under compression is found to be weaker than the effect of GB metastability on the yield or flow stresses for migration. A larger survey would be needed to assess the generality of this result. In the future, the athermal computations of yield and flow stress for GB migration can be extended to calculations of GB mobility at finite temperature for various irradiation induced metastable states. 

Finally, modeling efforts should be devoted to simulating high dose irradiation under more experimentally realistic irradiation conditions to analyze dynamic steady states which can form in the vicinity of pre-existing dislocations and grain boundaries as a function of radiation conditions such as temperature, dose rate, dose and ion type. The combination of atomistic simulation with in-situ experiments is expected to greatly improve our understanding of radiation induced dynamic steady states at the nanoscale. 

\section*{Conclusion}
\label{conclusion}

The CRA algorithm, previously applied to bulk materials, was used to examine the Frenkel pair accumulation response of a wide variety of GB and dislocation core structures in FCC Ni up to relatively large irradiation doses of 20 cDPA. The resulting athermal microstructural evolution was characterized for irradiated screw and edge dislocations, LAGBs with twist and tilt character and a variety of HAGBs. The mechanical responses of irradiated and pristine defect structures were compared under an applied shear strain. The following conclusions can be drawn from this work: 
\begin{enumerate}
\item Local steady state irradiation-induced local microstructures form in the vicinity of pre-existing topological defects at sufficiently high radiation doses. These microstructures depend on the initial defect character but not on the metastable structure (for the same average core density), have a higher energy than a pristine or randomly quenched microstructure, and are characterized by a balance between defect creation and annihilation processes. 
\item Examples of irradiated microstructures observed in this work include helical screw dislocations, jogged edge dislocations and GBs with roughened GB dislocation networks both in and out of the GB plane. Twist GBs are found to exhibit a higher propensity for roughening than symmetric tilt GBs. Athermal climb events are common, despite the overall conservation of mass of the CRA method. 
\item There is a correlation between the critical driving energy for saturation and pristine GB energy. On average, low energy GBs exhibit a larger increase in energy to reach a steady state than high energy GBs. 
\item The irradiation response of topological defects under CRA is intermediate between that of crystals and glasses, depending on the dislocation or GB character. In the high dose limit, glassy signatures of high energy GBs include exponential decay of relaxation spectra at steady state and correlated string-like displacements. The coherent twin GB embodies a crystalline response to irradiation with a power law decay of relaxation spectra at steady state and the formation of bulk dislocations at the GB, even within a narrow radiation zone.  
\item Dislocations must overcome a de-pinning stress to breakaway from their radiation zones under an applied shear strain, but recover the intrinsic mobilities of their pristine structures outside of the radiation zone. In contrast, many irradiated GBs are observed to exhibit softening or hardening effects compared to their pristine counterparts, even after migrating well outside of the radiation zone. These changes are attributed to a dependence of intrinsic GB mobility on metastable GB structure for the same nominal mode of migration. 
\end{enumerate}

\section*{Acknowledgements}

I.C. acknowledges support and computational resources from the LANL Advanced Simulation and Computing program and the LANL Postdoc Program via the Metropolis Fellowship. A.M., K.D, N.M., S.F. and A.H. acknowledge support from the U.S. Department of Energy through the LANL/LDRD Program, “Investigating How Material’s Interfaces and Dislocations Affect Strength” (iMIDAS) under grant no. 20210036DR. A.M. and N.M. also acknowledge the funding from the LANL/LDRD project 20220814PRD4: “Grain Boundary Characterization from Diffractograms by Physics-Informed Machine Learning.” B.P.U. acknowledges support as part of FUTURE (Fundamental Understanding of Transport Under Reactor Extremes), an Energy Frontier Research Center funded by the U.S. Department of Energy (DOE), Office of Science, Basic Energy Sciences (BES). The Los Alamos National Laboratory is operated by the Triad National Security, LLC, for the National Nuclear Security Administration of the U.S. Department of Energy (Contract No. 89233218CNA000001).

\bibliographystyle{elsarticle-num}
\bibliography{citations}

\section*{Figures and Tables} 

\newpage{}

\appendix
\global\long\def\tablename{Supplementary Table}%
\setcounter{figure}{0} \setcounter{table}{0} \setcounter{section}{0}

\setcounter{figure}{0}
 \let\oldthefigure\thefigure
 \renewcommand{\thefigure}{S\oldthefigure}

\end{document}


\begin{frontmatter}



\title{Supplementary information: the structure and migration of heavily irradiated grain boundaries and dislocations in Ni in the athermal limit}

\author[XCP5]{Ian Chesser\corref{mycorrespondingauthor}}
\author[PSI]{Peter M. Derlet}
\author[T1]{Avanish Mishra}
\author[MPA]{Sarah Paguaga}
\author[T1]{Nithin Mathew}
\author[MST]{Khanh Dang}
\author[MST]{Blas Pedro Uberuaga}
\author[XCP5]{Abigail Hunter}
\author[MPA]{Saryu Fensin}

\cortext[mycorrespondingauthor]{Corresponding author}

\address[XCP5]{XCP-5, Los Alamos National Laboratory, Los Alamos, New Mexico, NM USA}
\address[PSI]{Condensed Matter Theory Group, Paul Scherrer Institute, Villigen PSI, Switzerland}
\address[T1]{T-1, Los Alamos National Laboratory, Los Alamos, New Mexico, NM USA}
\address[MPA]{MPA-CINT, Los Alamos National Laboratory, Los Alamos, New Mexico, NM USA}
\address[MST]{MST-8, Los Alamos National Laboratory, Los Alamos, New Mexico, NM USA}


%


%
%
%

\end{frontmatter}

\global\long\def\tablename{Supplementary Table}%
\setcounter{figure}{0} \setcounter{table}{0} \setcounter{section}{0}

\setcounter{figure}{0}
 \let\oldthefigure\thefigure
 \renewcommand{\thefigure}{S\oldthefigure}

\section*{CRA applied to a Ni single crystal}

\begin{figure}[ht!]
\centering\leavevmode \includegraphics[width=0.9\textwidth]{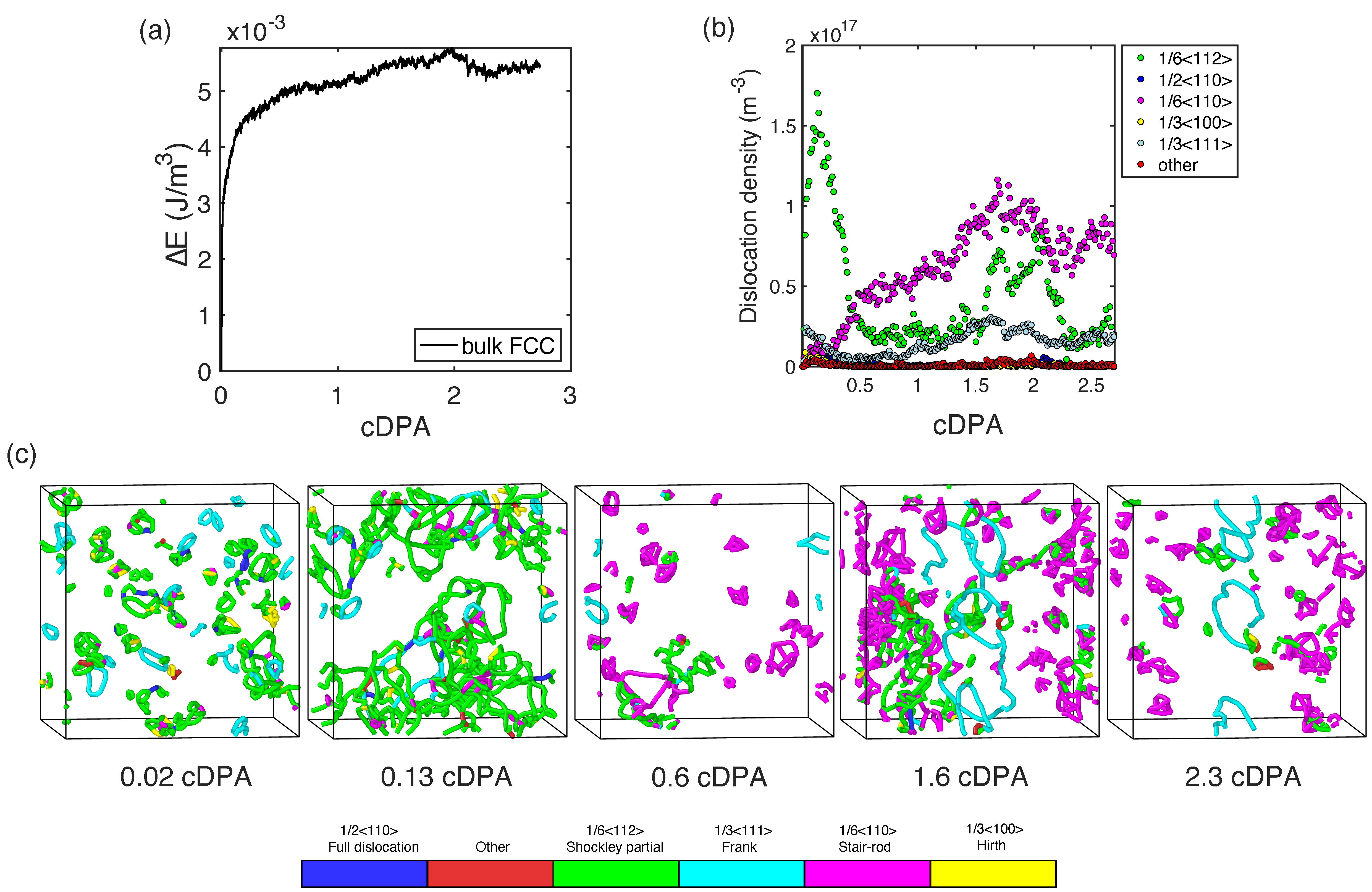}
\caption{CRA applied to bulk FCC Ni. Measured quantities include (a) excess energy and (b) dislocation density for different dislocation types. Snapshots of microstructural evolution at different doses are shown in (c). An extended network of Shockley partial dislocations formed by 0.13 cDPA is observed to collapse into an isolated distribution of truncated stacking fault tetrahedra with stair-rod character. Upon continued irradiation, an increase in density of stacking fault tetrahedra and Shockley partial dislocations is accompanied by the formation of a channel with primary Frank loop character. The steady state corresponds to dislocation migration and reaction events within this channeled microstructure.}
\label{fig:CRA_bulk}
\end{figure}

CRA was applied to single crystal Ni as a reference case for GB and dislocation simulations. Microstructural evolution is shown in Fig. \ref{fig:CRA_bulk}. Similar to prior work on BCC crystals \cite{CRA_BCC}, isolated dislocation loops are observed to nucleate at relatively small doses on the order of 0.01 cDPA and form system spanning dislocation  networks at larger doses on the order of 0.1 cDPA. Interestingly, in FCC Ni, the network structure reacts and collapses back to a state of predominantly isolated stacking fault tetrahedra by a dose of 0.6 cDPA. Upon continued irradiation, an increase in density of stacking fault tetrahedra and Shockley partial dislocations is accompanied by the formation of a channel with primary Frank loop character. An apparent steady state at doses of 2 cDPA and beyond corresponds to dislocation migration and reaction events within this channeled microstructure. 

\section*{Compression of pristine and irradiated bicrystals}

\begin{figure}[ht!]
\centering\leavevmode \includegraphics[width=0.8\textwidth]{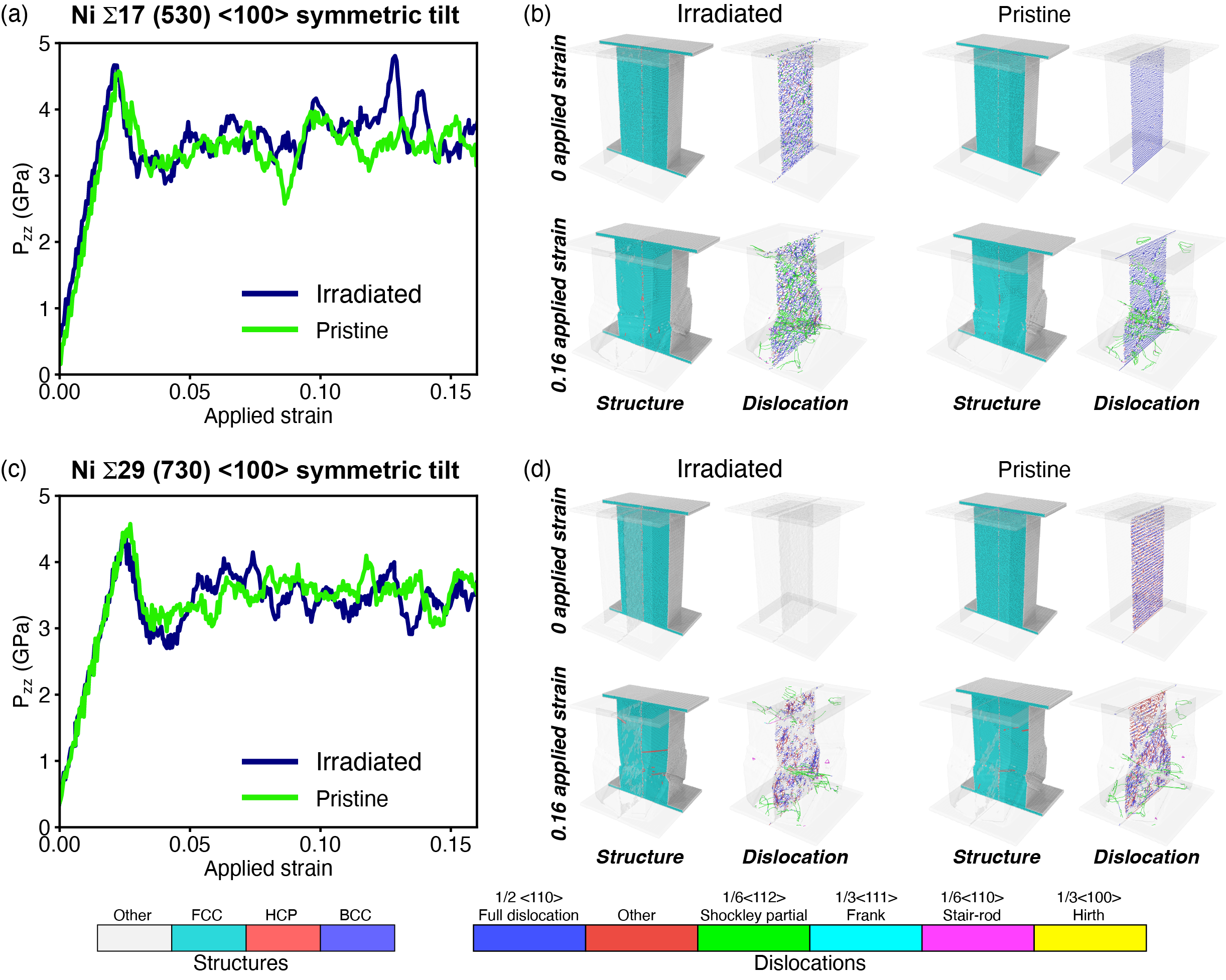}
\caption{Stress-strain responses of compressed Ni nanopillars, both irradiated and pristine, are presented for the (a) $\Sigma17 \hkl(530) \hkl<100>$ symmetric tilt GB and (c) $\Sigma 29 \hkl(730)\hkl<100>$ symmetric tilt GB. Corresponding dislocation microstructures are shown in (b) and (d), respectively. The atomistic snapshots have been sliced longitudinally to illustrate the deformation mechanisms more clearly}
\label{fig:Ni_nanopillar}
\end{figure}

\begin{figure}[ht!]
\centering\leavevmode \includegraphics[width=0.8\textwidth]{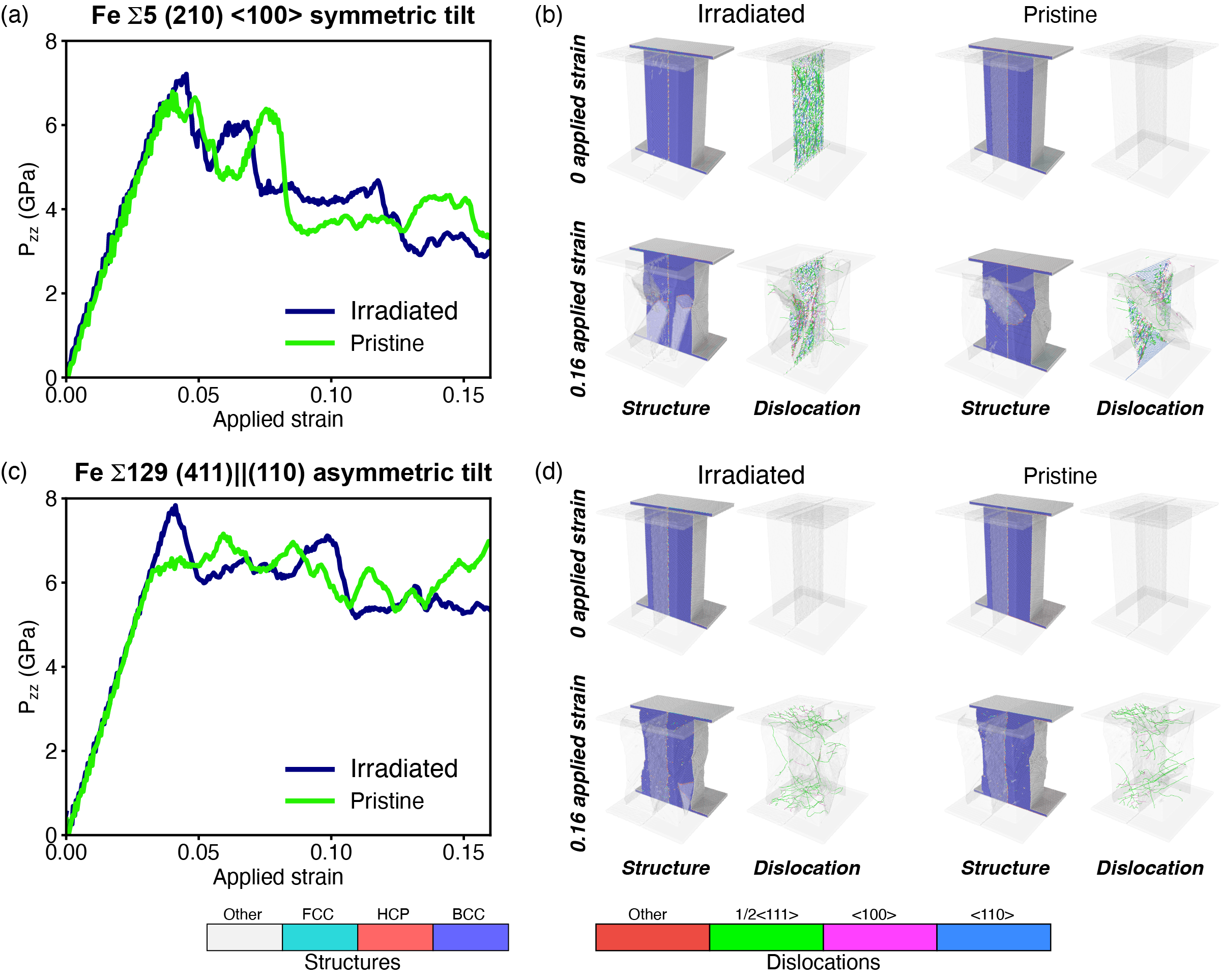}
\caption{Stress-strain responses of compressed Fe nanopillars, both irradiated and pristine, are presented for the (a) $\Sigma5 \hkl(210) \hkl<100>$ symmetric tilt GB and (c) $\Sigma 129 (411)||(110)$ asymmetric tilt GB. Corresponding dislocation microstructures are shown in (b) and (d), respectively. The atomistic snapshots have been sliced longitudinally to illustrate the deformation mechanisms more clearly}
\label{fig:Fe_nanopillar}
\end{figure}

As an additional test of the impact of radiation damage on mechanical properties, we performed nanopillar compression simulations with selected pristine and irradiated bicrystals in Ni and Fe using the method developed in \cite{mishra2023role}. The simulated nanopillars are illustrated in Fig. \ref{fig:Ni_nanopillar} and \ref{fig:Fe_nanopillar} and have relatively large dimensions with box sizes of 25 x 25 x 50 nm$^3$ and several million atoms each. GBs were positioned at the center of each bicrystal along the Z direction. Interatomic potentials were used consistent with those reported in the main text \cite{foiles2006computation,mendelev2003development}. Irradiated specimens were subjected to a total dose of 2 cDPA within a 1 nm radiation zone centered at each GB. Systems were irradiated with relatively small GB cross sections (around 5 nm x 5 nm) and replicated to form larger systems for subsequent nano-pillar compression tests. The examples of radiation induced hardening and softening for GB migration discussed in the main text (see Fig. 10) motivated our choice of \hkl<100> symmetric tilt GBs in Ni to test under compression. In Fe, one \hkl<100> symmetric tilt GB and asymmetric tilt GB were chosen for testing. 

Before compression, all nanopillars were equilibrated at a temperature of 300 K for a duration of 120 ps, employing the Nose-Hoover thermostat (NVT) and a timestep of 2 fs. Lateral boundaries perpendicular to the GB plane were set to shrink-wrapped (free) conditions, while a fixed boundary condition was maintained in the loading direction (Z). The deformation was applied by moving the top slab at a velocity of -0.07 $\AA$/ps, corresponding to a strain rate of approximately 10$^{8}$ s$^{-1}$. As the nanopillar underwent deformation, a central cuboid from the undeformed structure was chosen to evaluate the stress-strain behavior. The per-atom stress within this cuboid (refer to Figure S2 of the supplemental information in \cite{mishra2023role}) 
was aggregated and normalized by the total of the corresponding Voronoi volumes to determine the z-component of stress ($P_{zz}$). During the simulations, the structural changes in the atomistic snapshots were analyzed using polyhedral template matching \cite{larsen2016robust} and the dislocation extraction algorithm \cite{DXA} as implemented in OVITO \cite{OVITO}.

The impact of radiation induced GB metastability on the nanopillar compression responses of the chosen bicrystals was found to be relatively weak. Stress strain curves for pristine and irradiated samples in Fig. \ref{fig:Ni_nanopillar}a,c and \ref{fig:Fe_nanopillar}a,c exhibit similar yield and flow stresses with a larger effect of radiation damage on stress strain curves for Fe vs. Ni GBs. Both tilt GBs in Fe (one symmetric, one asymmetric) exhibit a transient radiation induced hardening effect upon yield. Simulation snapshots are shown for all systems at applied strains of 0 and 0.16 in Fig. \ref{fig:Ni_nanopillar}b,d and \ref{fig:Fe_nanopillar}b,d. At equivalent levels of deformation, the dislocation networks in the irradiated and pristine bicrystals are slightly different, with irradiated structures having distorted GB dislocation networks due to in-plane climb of edge dislocations. The distorted and reacted GB dislocations in the irradiated GB could be expected to influence its reactions with bulk dislocations compared to the ordered pristine GB. However, assessing the detailed impact of radiation damage on changes in GB-dislocation interactions and activated slip systems is beyond the scope of the current work. We suggest that, in order to observe a stronger effect of radiation induced metastability on the mechanical behavior of bicrystals in future work, larger irradiation zones and larger irradiated cross sections could be used to create steady state GB structures with significantly different morphologies and local dislocation environments compared to their pristine structures. Moreover,  only two \hkl<100> tilt GBs were tested for each material in this work, and future studies could survey a more diverse set of GBs. 

\newpage

\newpage

\bibliographystyle{elsarticle-num}
\bibliography{citations}